\documentclass{PoS}
\usepackage{amsmath}

\newcommand{\al}{\ensuremath{\alpha} }
\newcommand{\be}{\ensuremath{\beta} }
\newcommand{\ga}{\ensuremath{\gamma} }
\newcommand{\de}{\ensuremath{\delta} }
\newcommand{\eps}{\ensuremath{\epsilon} }
\newcommand{\la}{\ensuremath{\lambda} }
\newcommand{\La}{\ensuremath{\Lambda} }
\newcommand{\lalat}{\ensuremath{\lambda_{\rm lat}} }
\newcommand{\ka}{\ensuremath{\kappa} }
\newcommand{\Cbb}{\ensuremath{\mathbb C} }
\newcommand{\Ibb}{\ensuremath{\mathbb I} }
\newcommand{\Zbb}{\ensuremath{\mathbb Z} }
\newcommand{\cA}{\ensuremath{\mathcal A} }
\newcommand{\cAb}{\ensuremath{\overline{\mathcal A}} }
\newcommand{\cD}{\ensuremath{\mathcal D} }
\newcommand{\cDb}{\ensuremath{\overline{\mathcal D}} }
\newcommand{\cF}{\ensuremath{\mathcal F} }
\newcommand{\cFb}{\ensuremath{\overline{\mathcal F}} }
\newcommand{\cO}{\ensuremath{\mathcal O} }
\newcommand{\cN}{\ensuremath{\mathcal N} }
\newcommand{\cP}{\ensuremath{\mathcal P} }
\newcommand{\cQ}{\ensuremath{\mathcal Q} }
\newcommand{\cU}{\ensuremath{\mathcal U} }
\newcommand{\cUb}{\ensuremath{\overline{\mathcal U}} }
\newcommand{\cW}{\ensuremath{\mathcal W} }
\newcommand{\cWtw}{\ensuremath{\widetilde{\mathcal W}} }
\newcommand{\muhat}{\ensuremath{\widehat \mu} }
\newcommand{\glN}{\ensuremath{\mathfrak{gl}(N, \Cbb)} }
\newcommand{\X}{\ensuremath{\!\times\!} }
\newcommand{\pf}{\ensuremath{\mbox{pf}\,} }
\newcommand{\Tr}[1]{\ensuremath{\mbox{Tr}\left[ #1 \right]} }
\newcommand{\vev}[1]{\ensuremath{\left\langle #1 \right\rangle} }
\newcommand{\eq}[1]{Eq.~\ref{#1}}
\newcommand{\fig}[1]{Fig.~\ref{#1}}
\newcommand{\refcite}[1]{Ref.~\cite{#1}}
\newcommand{\secref}[1]{Section~\ref{#1}}
\def\figheight{7 cm}
\title{Results from lattice simulations \\ of $\cN = 4$ supersymmetric Yang--Mills}
\ShortTitle{Results from lattice simulations of $\cN = 4$ supersymmetric Yang--Mills}

\author{Simon Catterall$^{*\; a}$, Joel Giedt$^{*\; b}$, \speaker{David Schaich}$^{\ a}$, \hspace{6 cm} Poul H.~Damgaard$^c$ and Thomas DeGrand$^d$ \\
        $^a$ Department of Physics, Syracuse University, Syracuse, NY 13244, USA \\
        $^b$ Department of Physics, Applied Physics and Astronomy, \\ \ \ \ Rensselaer Polytechnic Institute, Troy NY 12065, USA \\
        $^c$ Niels Bohr International Academy and Discovery Center, Niels Bohr Institute, \\ \ \ \ University of Copenhagen, Copenhagen, Denmark \\
        $^d$ Department of Physics, University of Colorado, Boulder, CO 80309, USA}

\abstract{ % Draft complete
  We report recent results and developments from our ongoing lattice studies of $\cN = 4$ supersymmetric Yang--Mills theory.
  These include a proof that only a single fine-tuning needs to be performed, so long as the moduli space is not lifted by nonperturbative effects.
  We extend our investigations of supersymmetry restoration in the continuum limit by initiating Monte Carlo renormalization group studies.
  We present additional numerical evidence that the lattice theory does not suffer from a sign problem.
  Finally we study the static potential, which we find to be Coulombic at both weak and strong coupling.
  We compare the static potential Coulomb coefficients to perturbation theory, including initial results for $N = 3$ colors in addition to $N = 2$.
}

\FullConference{The 32nd International Symposium on Lattice Field Theory \\ 23--28 June 2014 \\ Columbia University, New York, NY}
\begin{document}
\setlength{\abovedisplayskip}{6 pt}
\setlength{\belowdisplayskip}{6 pt}
$\cN = 4$ supersymmetric Yang--Mills (SYM) is a fascinating quantum field theory which plays an important role in many areas of theoretical physics including holographic approaches to quantum gravity, understanding the structure of scattering amplitudes in $\cN = 8$ supergravity, and the conformal bootstrap program.
It is the only known example of a four-dimensional theory admitting a supersymmetric lattice discretization, and the only non-trivial example of a lattice gauge theory with a \be function that vanishes at least at one loop.
(The \be function of the continuum theory vanishes to all orders.)
Many people have contributed to the lattice formulation that we employ in our studies; see the review \cite{Catterall:2009it} and references therein for more information.
Alternate approaches to numerically studying $\cN = 4$ SYM include Refs.~\cite{Ishii:2008ib, Ishiki:2008te, Ishiki:2009sg, Hanada:2010kt, Honda:2011qk, Honda:2013nfa, Hanada:2013rga}.

In this proceedings we report recent results and developments from our ongoing lattice studies of $\cN = 4$ SYM.
Some of these results are new and preliminary; others have appeared in Refs.~\cite{Catterall:2011pd, Catterall:2012yq, Catterall:2013roa, Catterall:2014vka, Catterall:2014mha}.
In Sections~\ref{sec:twist}--\ref{sec:stab} we summarize the main elements of the lattice formulation.
In particular, in \secref{sec:Seff} we show that so long as the moduli space is not lifted by nonperturbative effects only a single marginal operator must be tuned to recover the $\cN = 4$ supersymmetric continuum limit.
Sections~\ref{sec:stab} and \ref{sec:susies} discuss observables that we measure to monitor supersymmetry restoration in the continuum limit, including initial results from new Monte Carlo renormalization group (MCRG) studies.
We briefly update our measurements of the Pfaffian phase in \secref{sec:pfaff}, exploring more lattice volumes for larger gauge groups.
The Pfaffian remains nearly real and positive in all cases, further evidence that the lattice theory does not suffer from a sign problem.
We focus on the static potential in \secref{sec:potential}, presenting initial results for $N = 3$ colors.
We conclude with some discussion of the next steps planned for our wide-ranging investigations.

To encourage independent work on lattice $\cN = 4$ SYM, we have developed a publicly available parallel software package suitable for use on clusters and supercomputers.\footnote{{\tt\href{http://github.com/daschaich/susy}{http://github.com/daschaich/susy}}}
This software evolved from the MILC code for lattice QCD, generalized to handle fermions in the adjoint representation for an arbitrary number of colors.
Its central feature is rational hybrid Monte Carlo importance sampling.
Many additional measurements are provided, including parallel computation of the complex Pfaffian of the fermion operator.
A detailed presentation recently appeared in \refcite{Schaich:2014pda}.
% ------------------------------------------------------------------

% ------------------------------------------------------------------
\section{\label{sec:twist}Discretizing the twisted continuum action} % Draft complete
The lattice theory we employ results from discretizing a topologically twisted form of continuum $\cN = 4$ SYM with gauge group SU($N$).
There are three independent topological twists of $\cN = 4$ SYM, of which we employ the Marcus or Geometric-Langlands twist~\cite{Marcus:1995mq, Kapustin:2006pk}.
This twisted theory is most conveniently written as a dimensional reduction of a five-dimensional action
\begin{equation}
  \label{caction}
  S = \frac{N}{2\la} \cQ \int d^5x \left(\chi_{ab}\cF_{ab} + \eta\left[\cDb_a, \cD_a\right] + \frac{1}{2}\eta d\right) - \frac{N}{8\la} \int d^5x\  \eps_{abcde}\ \chi_{ab}\cDb_c\chi_{de}.
\end{equation}
$\la = g^2 N$ is the 't~Hooft coupling.
The gauge fields $\cA_a$ (and their associated covariant derivatives $\cD_a$ and $\cDb_a$) have real parts which are just the usual Yang--Mills gauge fields, while their imaginary parts arise from the five scalars that appear after reducing $\cN = 1$ SYM in ten dimensions down to five dimensions.
The appearance of the scalar fields as vectors is odd at first sight, but is at the heart of the twisting process.

Twisting decomposes the fields under the diagonal subgroup of the Euclidean Lorentz symmetry and the R symmetry of the theory.
Since the scalars transform as vectors under the R symmetry they remain vectors under the twisted symmetry, and naturally combine with the R-singlet gauge fields.
Similar arguments show that twisting transforms the original four Majorana fermions into a set of antisymmetric fields with integer spins under the twisted symmetry, appearing as $(\eta, \psi_a, \chi_{ab})$ in the above action.
Likewise the $\cN = 4$ fermionic supercharges $Q_{\al i}$, $Q_{\dot\al i}^{\dag}$ become integer-spin $(\cQ, \cQ_a, \cQ_{ab})$.
In flat space twisting does not change the physical content of the theory.
It is merely an exotic change of variables, to a spinor-free form better suited to discretization.

The first part of the action in \eq{caction} is the \cQ variation of a function.
This scalar supercharge \cQ is the supersymmetry that can be preserved under discretization.
It acts on the fields as follows:
\begin{align}
  & \cQ\; \cA_a = \psi_a         & & \cQ\; \psi_a = 0                 \cr
  & \cQ\; \chi_{ab} = -\cFb_{ab} & & \cQ\; \cAb_a = 0 \label{eq:susy} \\
  & \cQ\; \eta = d               & & \cQ\; d = 0.     \nonumber
\end{align}
Its nilpotent character guarantees that the first piece of the action is trivially \cQ invariant.
The second term can be shown to be invariant using the Bianchi identity, so that $\cQ S = 0$.

We recover the standard Marcus twist of four-dimensional $\cN = 4$ SYM from naive dimensional reduction, by simply setting to zero the momentum in the fifth dimension, $\partial_4 = 0$, and identifying the corresponding component of the gauge field as the sixth scalar field, $\cA_4 \to \phi$.
A more symmetric choice is to project the theory onto the four-dimensional hyperplane with normal vector $\widehat{\mathbf n} = \frac{1}{\sqrt5}(1, 1, 1, 1, 1)$.
This corresponds to setting $\sum_{a = 0}^4 \partial_a = 0$ and $\sum_{a = 0}^4 \cA_a \to \phi$, preserving an $S_5$ permutation symmetry.

Using this formalism, the transition to the lattice is quite straightforward.
The lattice structure follows from a similar dimensional reduction.
Starting with the five-dimensional hypercubic lattice in momentum space, the constraint $\sum_{a = 0}^4 \partial_a = 0 \to \sum_{a = 0}^4 n_a = 0$ produces the $A_4$ lattice.
Transforming back to real space leaves us with the dual $A_4^*$ lattice, which retains an $S_5$ point group symmetry manifest as five (linearly dependent) basis vectors symmetrically spanning four space-time dimensions.
This $S_5$ symmetry provides a set of irreducible representations that match those of the continuum twisted SO(4) symmetry~\cite{Unsal:2006qp}.
It can be useful to think of $A_4^*$ as the four-dimensional analog of the triangular lattice $A_2^*$ in two dimensions, or as the weight lattice of SU(5).

The complex gauge fields $\cA_a(x)$ become complexified $\cU_a(n)$ living on the five links of the $A_4^*$ lattice.
The form of the nilpotent scalar supersymmetry \cQ remains the same as in the continuum (\eq{eq:susy}), and requires that these $\cU_a(n)$ be elements of the algebra $\glN$, with consequences we will discuss in \secref{sec:stab}.
Supersymmetry also forces us to assign the $\psi_a$ fermions to the same links, while the $\eta$ fermion and bosonic auxiliary field $d$ are placed on sites.
The $\chi_{ab}$ fermions are associated with the field strengths
\begin{equation*}
  \cFb_{ab}(n) = \cUb_a(n + \muhat_b) \cUb_b(n) - \cUb_b(n + \muhat_a) \cUb_a(n),
\end{equation*}
and connect lattice sites $n + \muhat_a + \muhat_b$ and $n$ so that their contractions with
\begin{equation*}
  \cD_a^{(+)} \psi_b = \cU_a(n)\psi_b(n + \muhat_a) - \psi_b(n)\cU_a(n + \muhat_b)
\end{equation*}
will produce gauge invariant loops.
After integrating out the auxiliary field $d$, on the lattice the two terms in \eq{caction} become
\begin{align}
  S_{exact}   & = \frac{N}{2\lalat} \sum_n a^4 \ \Tr{-\cFb_{ab}\cF_{ab}(n) + \frac{1}{2}\left(\cDb_a^{(-)}\cU_a(n)\right)^2 - \chi_{ab} \cD_{[a}^{(+)}\psi_{b]}^{\ }(n) - \eta \cDb^{(-)}_a\psi_a(n)} \nonumber         \\
  S_{closed}  & = -\frac{N}{8\lalat} \sum_n a^4 \ \Tr{\eps_{abcde}\ \chi_{de}(n + \muhat_a + \muhat_b + \muhat_c) \cDb^{(-)}_{c} \chi_{ab}(n)},                                                       \label{eq:Sexact}
\end{align}
where the specific form of the forward and backward difference operators can be found in \refcite{Catterall:2014vka}.

This exactly supersymmetric lattice action was studied analytically in Refs.~\cite{Catterall:2011pd, Catterall:2013roa, Catterall:2014mha}.
These studies have found that the single exact supersymmetry suffices to establish the following results:
%\vspace{-6 pt}
\begin{itemize}
  \item The moduli space survives to all orders of lattice perturbation theory, which prohibits any scalar potential being induced through radiative corrections.
  \item The \be function vanishes at least at one loop in lattice perturbation theory.
  \item Certain quantities, such as the partition function, may be computed exactly in the semi-classical limit.
  \item Lattice symmetries imply that only a single tuning of a marginal operator is necessary to restore R symmetry and all supersymmetries in the continuum limit.
\end{itemize}
We derive the last result in the next section.
% ------------------------------------------------------------------

% ------------------------------------------------------------------
\section{\label{sec:Seff}Renormalization and the long-distance effective theory} % Draft complete
Since $\cN = 4$ SYM possesses a line of conformal fixed points, we reach the continuum limit of the lattice theory by taking $1 / L \to 0$ for any fixed 't~Hooft coupling $\la$, where $L$ is the linear length of the lattice volume.
We must ensure that the long-distance effective action $S_{eff}$ of the lattice theory appropriately recovers $\cN = 4$ SYM in the continuum limit.
Simply defining $S_{eff}$ requires the existence of a real-space renormalization group (RG) blocking transformation that preserves both the symmetries of the system and the geometric interpretation of the fields.
In this section we will present an explicit example of such a blocking scheme, derive the long-distance effective action, and show that only a single tuning is required if the moduli space is not lifted by nonperturbative effects.
These results were recently derived in \refcite{Catterall:2014mha}.

Writing the original $A_4^*$ lattice as $\La = \left\{a \sum_{i = 0}^3 n_i \muhat_i \ | \ n \in \Zbb^4\right\}$, with $\muhat_i$ the first four of the five (degenerate) basis vectors, the blocked lattice $\La' = \left\{2a \sum_{i = 0}^3 n_i \muhat_i \ | \ n \in \Zbb^4\right\}$ is merely doubled in every direction.
The blocked fields must begin and end on sites of $\La'$.
We will denote the blocked fields by primes, and work in lattice units with $a = 1$.
The following blocking transformation preserves both the \cQ algebra and the geometric interpretation of the fields:
\begin{align}
  \cU_a'(n)     = & \xi \cU_a(n) \cU_a(n + \muhat_a)    \hspace{4 cm} \cUb_a'(n)  = \xi \cUb_a(n + \muhat_a) \cUb_a(n)                                                                            \nonumber         \\
  d'(n)         = & d(n)                                \hspace{7 cm} \eta'(n)    = \eta(n)                                                                                                       \nonumber         \\
  \psi_a'(n)    = & \xi \left[\psi_a(n) \cU_a(n + \muhat_a) + \cU_a(n) \psi_a(n + \muhat_a)\right]                                                                                                \label{eq:block}  \\
  \chi_{ab}'(n) = & \frac{\xi^2}{2} \left[\cUb_a(n + \muhat_a + 2\muhat_b) \cUb_b(n + \muhat_a + \muhat_b) + \cUb_b(n + 2\muhat_a + \muhat_b) \cUb_a(n + \muhat_a + \muhat_b)\right] \chi_{ab}(n) \nonumber         \\
                  & + \xi^2 \left[\cUb_a(n + \muhat_a + 2\muhat_b) \chi_{ab}(n  + \muhat_b) \cUb_b(n) + \cUb_b(n + 2\muhat_a + \muhat_b) \chi_{ab}(n  + \muhat_a) \cUb_a(n)\right]                \nonumber         \\
                  & + \frac{\xi^2}{2} \chi_{ab}(n + \muhat_a + \muhat_b)\left[\cUb_a(n + \muhat_b) \cUb_b(n) + \cUb_b(n + \muhat_a) \cUb_a(n)\right].                                             \nonumber
\end{align}
Because the link variables are non-compact elements of $\glN$, we allow for the possibility that they are rescaled by a factor $\xi$ under the transformation.
For the site variables $\eta$ and $d$ we simply use decimation.
From \eq{eq:block} it is easy to see that $\cU_a'$ and $\psi_a'$ connect lattice sites $n$ and $n + 2\muhat_a$, while $\chi_{ab}'$ connects $n + 2\muhat_a + 2\muhat_b$ and $n$, as desired.
As a consequence, the properties of the system under the $S_5$ point group symmetry are preserved.
Any $S_5$ invariant of the original fields, such as $\sum_a \cU_a \cUb_a$, remains invariant when expressed in terms of the blocked fields.

It is obvious that $\cQ \cUb_a' = 0$, $\cQ \eta' = d'$ and $\cQ d' = 0$, just as in \eq{eq:susy}.
$\cQ \cU_a' = \psi_a'$ also follows from the original algebra, while $\cQ \psi_a' = 0$ due to the negative sign from anticommuting \cQ past $\psi_a$:
\begin{equation*}
    \cQ \psi_a'(n) = \xi \left[-\psi_a(n) \psi_a(n + \muhat_a) + \psi_a(n) \psi_a(n + \muhat_a)\right] = 0.
\end{equation*}
Finally, defining the blocked field strength as expected,
\begin{equation*}
  \cFb_{ab}'(n) = \cUb_a'(n + \muhat_b) \cUb_b'(n) - \cUb_b'(n + \muhat_a) \cUb_a'(n),
\end{equation*}
just a few lines of algebra are required to check $\cQ \chi_{ab}' = \cF_{ab}'$.
So we see that the blocking scheme in \eq{eq:block} preserves the supersymmetry $\cQ^2 = 0$ along with the other symmetries of the system.

We can now consider what lattice operators could possibly be generated under RG flow based on this blocking scheme.
Those that give relevant or marginal operators in the continuum limit must be included in the most general long-distance effective action $S_{eff}$.
If two lattice operators produce the same relevant or marginal continuum operator, and only differ by irrelevant operators in the continuum limit, then we only need to include one of them in $S_{eff}$.
We must ensure that every operator in the effective action acquires its canonical coefficient, either by fine-tuning the corresponding coefficients in the UV theory, \eq{eq:Sexact}, or by adding new counterterms to the lattice action.
As we will describe below, some operators can be given their canonical coefficients simply by rescaling the fields.

In both the continuum and lattice theories, there is only one renormalizable $\cQ$-closed operator: the second line of \eq{eq:Sexact} gives the second term in \eq{caction}.
The only renormalizable $\cQ$-exact terms must take the form $\cQ \Tr{\Psi f(\cU, \cUb, d)}$ or $\cQ \left\{\Tr{\eta} \Tr{f(\cU, \cUb, d)}\right\}$ where $\Psi$ stands for one of the fermion fields.
\cQ must act on a fermionic quantity so that $S_{eff}$ is bosonic, while cubic or higher powers of fermions would be nonrenormalizable.
Only $\eta$ can be used in double-trace operators, since traces must be evaluated at sites to be gauge invariant.
Finally, the effective action must be invariant under two further global symmetries of the system: the shift symmetry
\begin{equation}
  \eta \to \eta + c\Ibb_N,
\end{equation}
with $c$ a constant Grassmann parameter, and a U(1) ``ghost number'' symmetry~\cite{Catterall:2009it} corresponding to the untwisted part of the SO(6) R symmetry.

For $\Psi = \eta$ these requirements leave only three possible terms,
\begin{align}
  & \cQ \Tr{\eta \cDb_a^{(-)} \cU_a} &
  & \cQ \Tr{\eta d} &
  & \cQ \Tr{\eta \cU_a \cUb_a} - \frac{1}{N}\cQ \left\{\Tr{\eta} \Tr{\cU_a \cUb_a}\right\},
\end{align}
the first two of which are already present in \eq{eq:Sexact}.
The only nontrivial operator with $\Psi = \psi_a$ is forbidden by the U(1) ghost number symmetry~\cite{Catterall:2014mha}.
Finally, for $\Psi = \chi_{ab}$ the antisymmetry of $\chi_{ab}$ requires that $\cQ \Tr{\chi_{ab} \cU_a \cU_b}$ and $\cQ \Tr{\chi_{ab} \cU_b \cU_a}$ be combined with a negative sign, leaving
\begin{equation}
  \cQ \Tr{\chi_{ab} \cF_{ab}}
\end{equation}
which is also already in \eq{eq:Sexact}.
These arguments only hold because the blocking scheme in \eq{eq:block} preserves the geometric interpretation of the blocked fields on $\La'$, so that the same terms appear in both the original action and the long-distance effective theory.
It is also important that the blocking preserves the $S_5$ symmetry, without which many other operators would have been generated.

Thus the most general long-distance effective action is
\begin{equation}
  \begin{split}
    S_{eff} & = \cQ \Tr{\al_1 \chi_{ab} \cF_{ab} + \al_2 \eta \cDb_a^{(-)} \cU_a - \frac{\al_3}{2} \eta d} - \frac{\al_4}{4} \eps_{abcde} \Tr{\chi_{de} \cDb_c^{(-)} \chi_{ab}} \\
    & \qquad + \be \cQ \left\{ \Tr{\eta \cU_a \cUb_a} - \frac{1}{N} \Tr{\eta} \Tr{\cU_a \cUb_a} \right\},                                                                       \label{eq:Seff0}
  \end{split}
\end{equation}
suppressing the overall $\frac{N}{2\lalat} \sum_n a^4$ for brevity.
Acting with $\cQ$, followed by the rescaling
\begin{align}
  \eta      & \to \la_{\eta} \eta       &
  \psi_a    & \to \la_{\psi} \psi_a     &
  \chi_{ab} & \to \la_{\chi} \chi_{ab}  &
  d         & \to \la_d d,   \label{eq:rescale}
\end{align}
%gives
\begin{align}
  S_{eff} & = \Tr{-\al_1 \cFb_{ab} \cF_{ab} - \al_1 \la_{\chi} \la_{\psi} \chi_{ab} \cD_{[a}^{(+)}\psi_{b]}^{\ } + \al_2 \la_d d \cDb_a^{(-)} \cU_a - \al_2 \la_{\eta} \la_{\psi} \eta \cDb_a^{(-)} \psi_a - \frac{\al_3}{2} \la_d^2 d^2} \nonumber \\
  & - \frac{\al_4}{4} \la_{\chi}^2 \eps_{abcde} \Tr{\chi_{de} \cDb_c^{(-)} \chi_{ab}}                                                                                                                                                               \\
  & + \be \left\{ \la_d \Tr{d \cU_a \cUb_a} - \la_{\eta} \la_{\psi} \Tr{\eta \psi_a \cUb_a} - \frac{\la_d}{N} \Tr{d} \Tr{\cU_a \cUb_a} + \frac{\la_{\eta} \la_{\psi}}{N} \Tr{\eta} \Tr{\psi_a \cUb_a} \right\}.                           \nonumber
\end{align}
We are now free to set many of the coefficients to $\al_1$ by imposing four constraints,
\begin{align}
  \al_1 \la_{\chi} \la_{\psi} & = \al_1 &
  \al_2 \la_d                 & = \al_1 &
  \al_2 \la_{\eta} \la_{\psi} & = \al_1 &
  \al_4 \la_{\chi}^2          & = \al_1.
\end{align}
Solving this system of equations produces the rescaling
\begin{align}
  \la_{\eta} & = \sqrt{\frac{\al_1^3}{\al_4 \al_2^2}}               &
  \la_{\chi} & = \frac{1}{\la_{\psi}} = \sqrt{\frac{\al_1}{\al_4}}  &
  \la_d      & = \frac{\al_1}{\al_2}.
\end{align}
We can then absorb an overall factor of $\al_1$ into the renormalized gauge coupling, which does not need to be tuned since the continuum theory is conformal for any $\la$.
This leaves
\begin{align}
  S_{eff} & = \Tr{-\cFb_{ab} \cF_{ab} - \chi_{ab} \cD_{[a}^{(+)}\psi_{b]}^{\ } + d \cDb_a^{(-)} \cU_a - \eta \cDb_a^{(-)} \psi_a - \frac{1}{2} \frac{\al_1 \al_3}{\al_2^2} d^2}   \nonumber       \\
  & \qquad - \frac{1}{4} \eps_{abcde} \Tr{\chi_{de} \cDb_c^{(-)} \chi_{ab}}                                                                                                       \label{eq:Seff} \\
  & \qquad + \frac{\be}{\al_2} \left\{ \Tr{d \cU_a \cUb_a} - \Tr{\eta \psi_a \cUb_a} - \frac{1}{N} \Tr{d} \Tr{\cU_a \cUb_a} + \frac{1}{N} \Tr{\eta} \Tr{\psi_a \cUb_a} \right\}.  \nonumber
\end{align}

At most two fine-tunings are required to obtain continuum $\cN = 4$ SYM: $\al_3 \to \al_2^2 / \al_1$ and $\be \to 0$.
In particular, it is remarkable that the relation $\la_d = \la_{\eta} \la_{\psi}$ keeps the \be term from bifurcating into multiple operators with different coefficients.
We can already appreciate the benefits of maintaining an exact supersymmetry at nonzero lattice spacing by comparing these two tunings against the eight that would be necessary with naive Wilson fermions~\cite{Catterall:2014mha}.
In addition, as mentioned at the end of \secref{sec:twist}, this exact supersymmetry ensures that the moduli space survives to all orders of lattice perturbation theory~\cite{Catterall:2011pd}.
\refcite{Catterall:2014mha} demonstrates that the \be term in \eq{eq:Seff} lifts the moduli space.
Therefore if nonperturbative effects such as instantons also preserve the moduli space, then the \be term is forbidden and $\be = 0$ is preserved by the RG flow.

In the absence of the \be term, we can easily integrate out the auxiliary field $d$ in \eq{eq:Seff}.
The resulting long-distance effective action reproduces \eq{eq:Sexact} up to one change:
\begin{equation}
  \label{eq:c2}
  \frac{1}{2}\left(\cDb_a^{(-)}\cU_a(n)\right)^2 \longrightarrow \frac{c_2}{2}\left(\cDb_a^{(-)}\cU_a(n)\right)^2
\end{equation}
in $S_{exact}$, with $c_2 = 2 - \al_2^2 / (\al_1 \al_3)$.
That is, a single fine-tuning of this marginal $c_2$ should suffice to recover $\cN = 4$ SYM in the continuum limit defined by $1 / L \to 0$ with fixed $\lalat$.
We have begun to explore this tuning in numerical calculations.
Before we describe our initial results in \secref{sec:susies}, we must discuss certain sources of soft \cQ supersymmetry breaking that are required to stabilize numerical simulations of lattice $\cN = 4$ SYM.
% ------------------------------------------------------------------

% ------------------------------------------------------------------
\section{\label{sec:stab}Stabilizing numerical computations} % Draft complete
\begin{figure}[bp]
  \centering
  \includegraphics[height=\figheight]{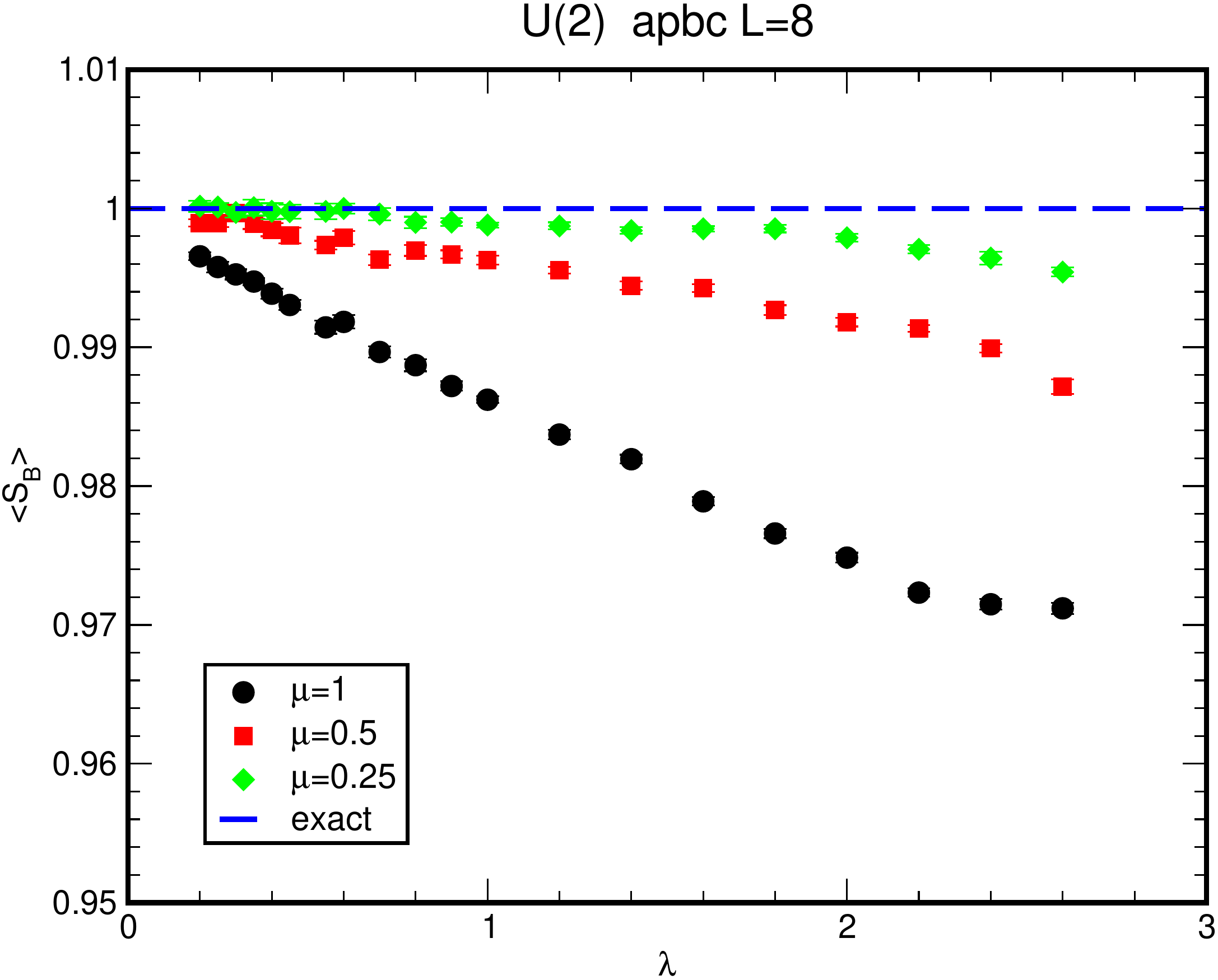}
\caption{\label{fig:sB} Relative deviations of the bosonic action from its exact supersymmetric value, $\vev{s_B} / 18$, plotted vs.\ the 't~Hooft coupling \lalat for $8^4$ lattices with several nonzero values of the $\cQ$-breaking coefficient $\mu$ in \protect\eq{eq:mu}.  As $\mu \to 0$, $\vev{s_B} / 18 \to 1$ for all $\lalat$, indicating the restoration of supersymmetry.  From \protect\refcite{Catterall:2012yq}.}
\end{figure}

For the lattice theory to recover $\cN = 4$ SYM in the continuum limit, we must also require that the lattice gauge fields have the expansion
\begin{equation}
  \label{eq:vev}
  \cU_a(n) = \frac{1}{a}\Ibb_N + \cA_a(x) + \cO(a)
\end{equation}
in some appropriate gauge.
Without the unit matrix the lattice $\cF_{ab}$ would lack the derivative (kinetic) terms required to correspond to the continuum field strength.
If the link variables were elements of the gauge group this unit matrix would arise automatically from expanding the exponential of terms in the algebra.
\eq{eq:vev} is nontrivial since our links are themselves elements of the algebra $\glN$, which also implies that the lattice gauge group is U($N$) as opposed to the target SU($N$)~\cite{Palumbo:1990kh, Becchi:1992um, Palumbo:2001br}.
In the continuum limit the U(1) degrees of freedom decouple as $\mbox{U}(N) = \mbox{SU}(N)\otimes \mbox{U}(1)$, and below we will discuss how to suppress lattice artifacts associated with the U(1) sector.
In addition, the integration measure for the lattice gauge fields is not the usual Haar measure but a flat measure.
This flat measure is gauge invariant, since the fields are complex and the measure contains both $D\cU$ and $D\cUb$.
The Jacobian resulting from a gauge transformation on $D\cU$ cancels against the corresponding quantity for $D\cUb$.

The appearance of the unit matrix in \eq{eq:vev} corresponds to the U(1) component of the scalar fields taking on a vacuum expectation value.
While this is indeed a classical vacuum state it is not unique.
There are infinitely many flat directions in the theory and a priori it is not clear that the vacuum needed to generate the kinetic terms is picked out.
To stabilize this vacuum we include in the lattice action (with no sum over repeated indices)
\begin{equation}
  \label{eq:mu}
  \de S_1 = \frac{N}{2\la}\mu^2 \sum_{n,\ c} a^4 \left(\frac{1}{N}\Tr{\cU_c(n) \cUb_c(n)} - 1\right)^2.
\end{equation}
This term also lifts the SU($N$) flat directions, biasing the system towards the continuum superconformal phase in which the scalars do not pick up any non-trivial expectation values.

\begin{figure}[bp]
  \centering
  \includegraphics[height=\figheight]{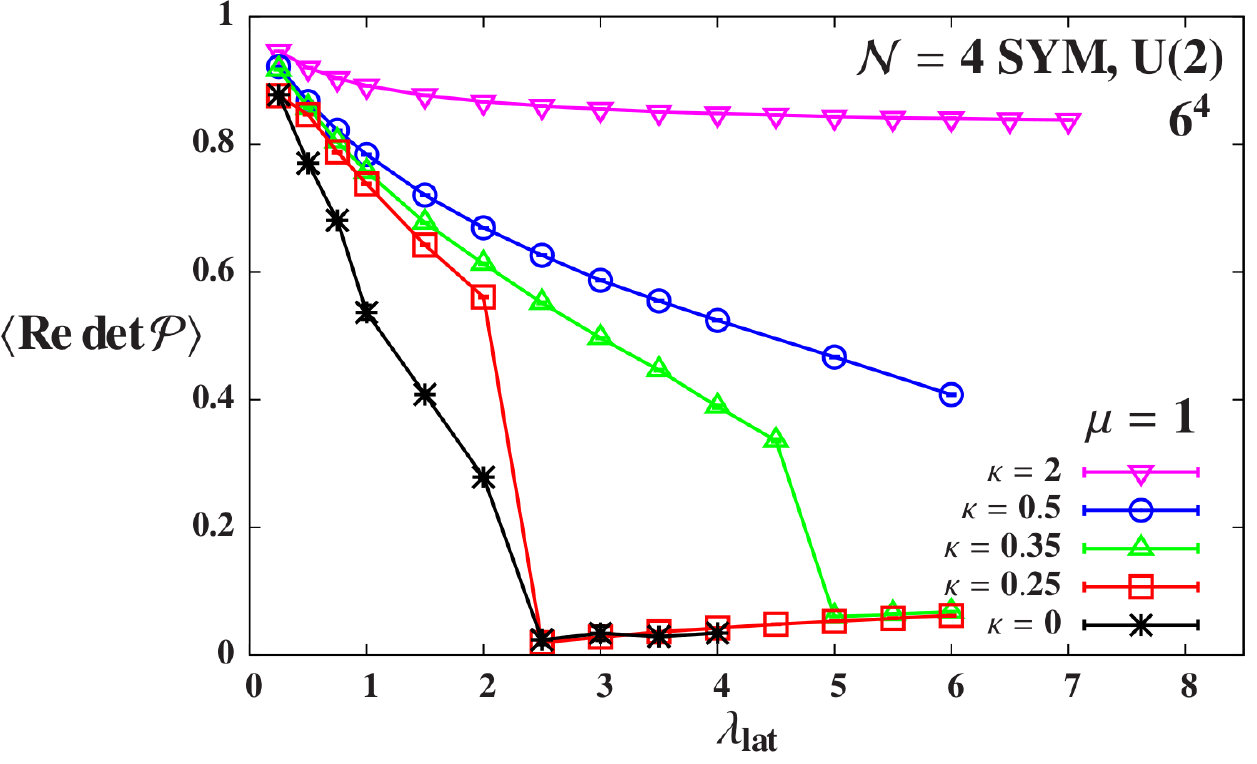}
  \caption{\label{fig:plaq_det} The real part of the plaquette determinant vs.\ the 't~Hooft coupling \lalat on $6^4$ lattices with a variety of \ka in \protect\eq{eq:ka} and fixed $\mu = 1$.  As \ka increases the confinement transition associated with the U(1) sector moves to larger $\lalat$, and disappears entirely for $\kappa \geq 0.5$.  Lines connect points with the same \ka to guide the eye.  From \protect\refcite{Catterall:2014vka}.}
\end{figure}

Although $\de S_1$ breaks supersymmetry, it does so softly: the exact \cQ supersymmetry at $\mu = 0$ guarantees that any $\cQ$-breaking counterterms must possess coefficients that vanish as $\mu \to 0$.
This can be confirmed by considering the bosonic action per lattice site, $\vev{s_B}$.
This quantity is related to a \cQ Ward identity, which for unbroken \cQ supersymmetry predicts the exact $\la$-independent value $s_B = 9N^2 / 2$ for gauge group U($N$).
In \fig{fig:sB} we plot $\vev{s_B} / 18$ (for $N = 2$) vs.\ the 't~Hooft coupling for several nonzero values of $\mu$.
Clearly as $\mu \to 0$ we approach the supersymmetric value for all 't~Hooft couplings.

We mentioned above the issue of lattice artifacts linked to the undesired U(1) sector.
The determinant of the plaquette is a gauge-invariant quantity associated with this sector.
For sufficiently large $\lalat \approx 2.5$, in \refcite{Catterall:2014vka} we observed $\vev{\mbox{Re}\det \cP}$ falling to zero as shown by the black bursts in \fig{fig:plaq_det}.
At the same \lalat the system confined, with the Polyakov loop also vanishing while the fermion operator developed a large number of near-zero eigenvalues.
While \fig{fig:plaq_det} considers $N = 2$, we have also observed all the same effects with gauge group U(1).

This behavior suggests that we are observing confinement in the compact lattice U(1) sector, which has a well-known dual description in terms of monopole world lines, one-dimensional objects that form closed loops of monopole flux.
Indeed, if we measure the density of monopole world lines we find it becomes nonzero at precisely the same $\lalat \approx 2.5$~\cite{Catterall:2014vka}.
To suppress this lattice artifact we penalize small plaquette determinant values by adding to the lattice action
\begin{equation}
  \label{eq:ka}
  \de S_2 = \ka \sum_{n,\ a < b} a^4\ |\det \cP_{ab}(n) - 1|^2,
\end{equation}
where $\cP_{ab}$ is the plaquette in the $a$--$b$ plane.
To leading order in the lattice spacing, $\de S_2$ just generates a U(1) field strength term
\begin{equation*}
  \de S_2 = 2\ka \sum_{n,\ a < b} a^4 \left[1 - \cos{\cF_{ab}^0(n)}\right] + \cO(a^5).
\end{equation*}
In agreement with simulations of compact U(1) gauge theory~\cite{DeGrand:1980eq}, we observe that for $\ka \geq 0.5$ we suppress monopoles and avoid the confinement transition for arbitrarily large 't~Hooft coupling, as shown in \fig{fig:plaq_det}.

\begin{figure}[bp]
  \includegraphics[width=0.45\linewidth]{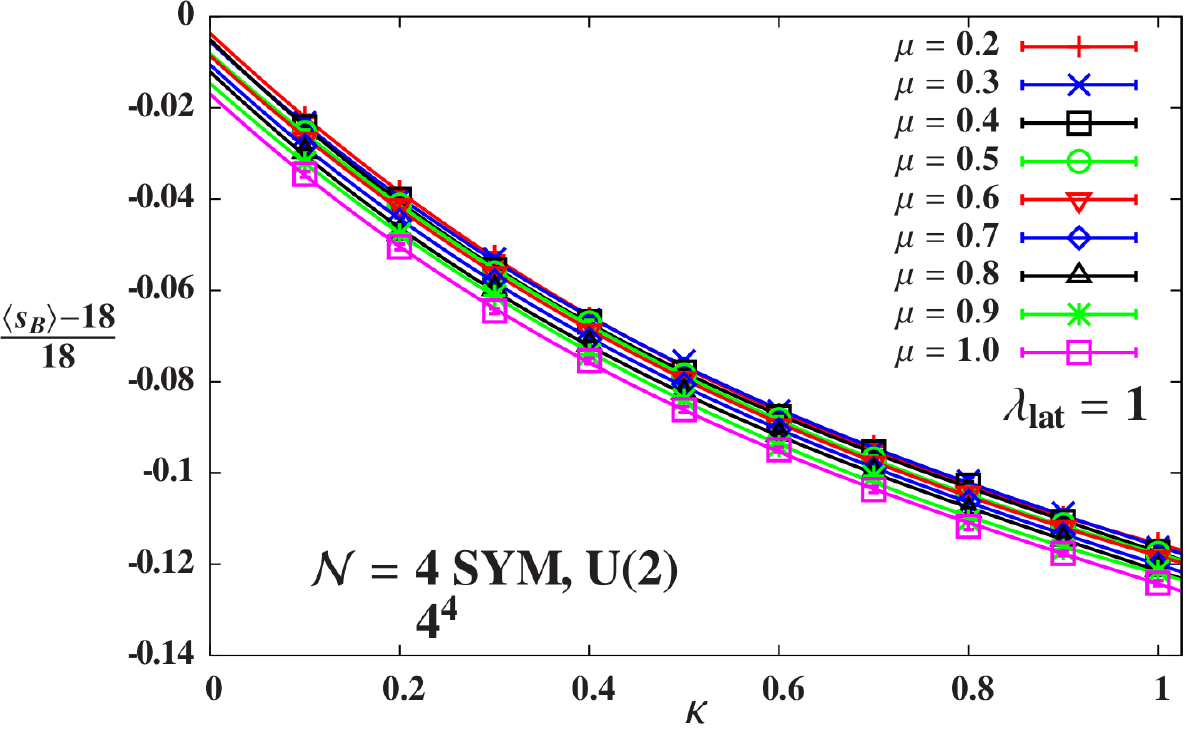}\hfill \includegraphics[width=0.45\linewidth]{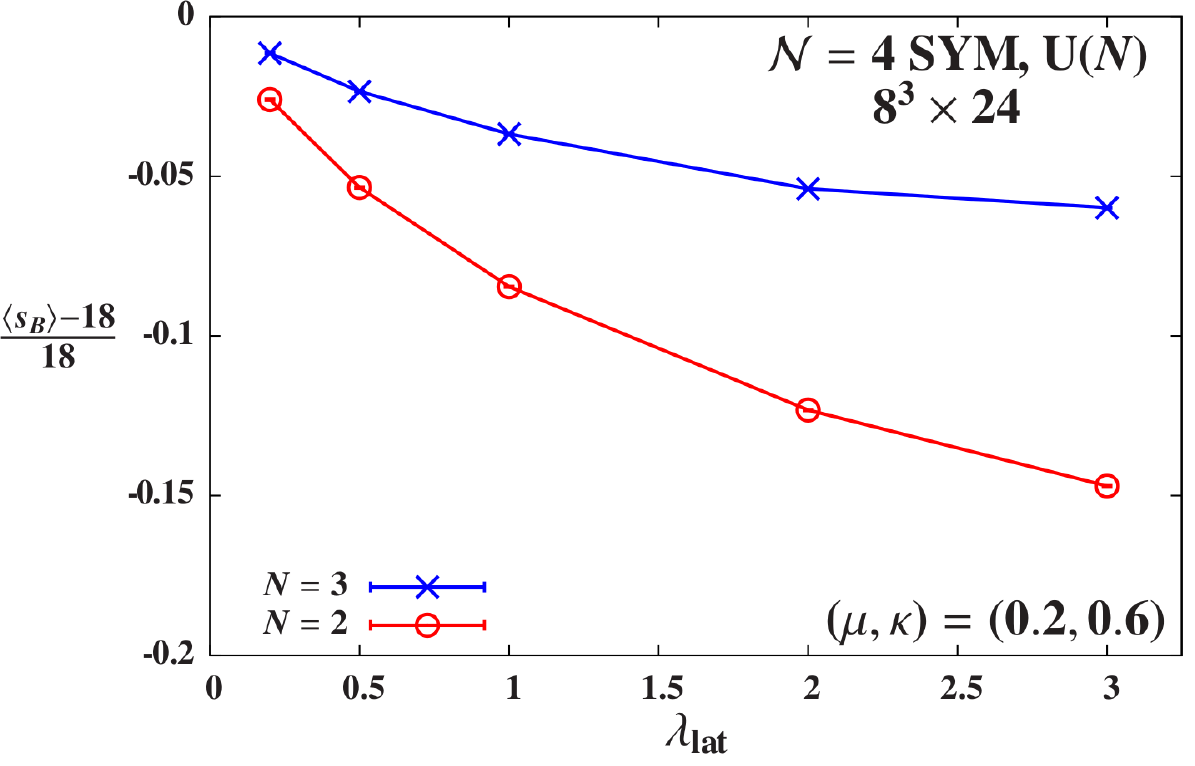}
  \caption{\label{fig:sB_ka} Relative deviations of the bosonic action from its exact supersymmetric value.  {\bf Left} (from \protect\refcite{Catterall:2014vka}): We fix $\lalat = 1$ and plot vs.\ \ka for $4^4$ lattices with several nonzero $\mu$.  The lines are cubic $\ka \to 0$ extrapolations that confirm the restoration of supersymmetry in the limit $(\mu, \ka) \to (0, 0)$.  {\bf Right:} We fix $(\mu, \ka) = (0.2, 0.6)$ and plot vs.\ \lalat for $8^3\X 24$ lattices with gauge groups U(2) and U(3), finding $\sim$10\% deviations that decrease $\propto 1 / N^2$.  The lines connect points to guide the eye.}
\end{figure}

The lattice action we use in numerical computations includes all of Eqs.~\ref{eq:Sexact}, \ref{eq:mu} and \ref{eq:ka}:
\begin{equation}
  S = S_{exact} + S_{closed} + \de S_1 + \de S_2.
\end{equation}
Both $\de S_1$ and $\de S_2$ softly break the \cQ supersymmetry.
As in \fig{fig:sB}, the left panel of \fig{fig:sB_ka} shows that the bosonic action $s_B$ acquires its exact supersymmetric value in the limit $(\mu, \ka) \to (0, 0)$, indicating the recovery of supersymmetry.
This plot also makes it clear that nonzero \ka leads to much more severe supersymmetry breaking than does nonzero $\mu$.
However, this effect is confined to the U(1) sector that decouples in the continuum limit.
The right panel of \fig{fig:sB_ka} illustrates the $\sim$10\% level of \cQ supersymmetry breaking on representative $8^3\X 24$ lattice ensembles that we will discuss further below.
Here we compare $N = 2$ and 3, finding that violations of supersymmetry decrease $\propto 1 / N^2$.
In \refcite{Catterall:2014vka} we discuss the breaking and restoration of \cQ in more detail.
% ------------------------------------------------------------------

% ------------------------------------------------------------------
\section{\label{sec:susies}Restoration of the other 15 supersymmetries $\cQ_a$ and $\cQ_{ab}$} % Draft complete
Although the lattice formulation discussed above exactly preserves the \cQ supersymmetry for $(\mu, \ka) \to (0, 0)$, the other fifteen $\cQ_a$ and $\cQ_{ab}$ are broken, and must be recovered in the continuum limit.
In \refcite{Catterall:2013roa} we showed how restoration of the full symmetries of $\cN = 4$ SYM follows from preservation of both \cQ and any one of a set of discrete R symmetries $\left\{R_a, R_{ab}\right\}$, subgroups of the continuum SO(6)$_R$ symmetry that fix the correct coefficients in the long-distance effective action, \eq{eq:Seff0}.
On the lattice, the $R_a$ acts on the links as~\cite{Catterall:2014vka}
\begin{align}
  \label{eq:Rtrans}
  & R_a \cU_a  = \cU_a        &
  & R_a \cUb_a = \cUb_a       &
  & R_a \cU_b  = \cUb_b^{-1}  &
  & R_a \cUb_b = \cU_b^{-1},
\end{align}
for all $b \ne a$.
This transformation commutes with lattice gauge invariance, allowing us to measure violations of the $R_a$ symmetry by considering its action on $m\X n$ Wilson loops in the $a$--$b$ plane,
\begin{equation}
  \begin{split}
    \cW_{ab} & = \Tr{\prod_m \cU_a(x) \prod_n \cU_b(x + m\muhat_a) \prod_m \cUb_a(x + n\muhat_b) \prod_n \cUb_b(x)}                                           \\
    \implies R_a \cW_{ab} & = \Tr{\prod_m \cU_a(x) \prod_n \cUb_b^{-1}(x + m\muhat_a) \prod_m \cUb_a(x + n\muhat_b) \prod_n \cU_b^{-1}(x)} \equiv \cWtw_{ab}  \label{eq:Rsymm}.
  \end{split}
\end{equation}
Since our links are non-unitary, $\cWtw_{ab} \ne \cW_{ab}$ even though they follow the same path in the lattice.
By computing the relative difference $(\cWtw - \cW) / \frac{1}{2}(\cWtw + \cW)$ we can assess how badly $R_a$ is broken on the lattice and monitor its restoration as we approach the continuum limit.

\begin{figure}[bp]
  \includegraphics[width=0.45\linewidth]{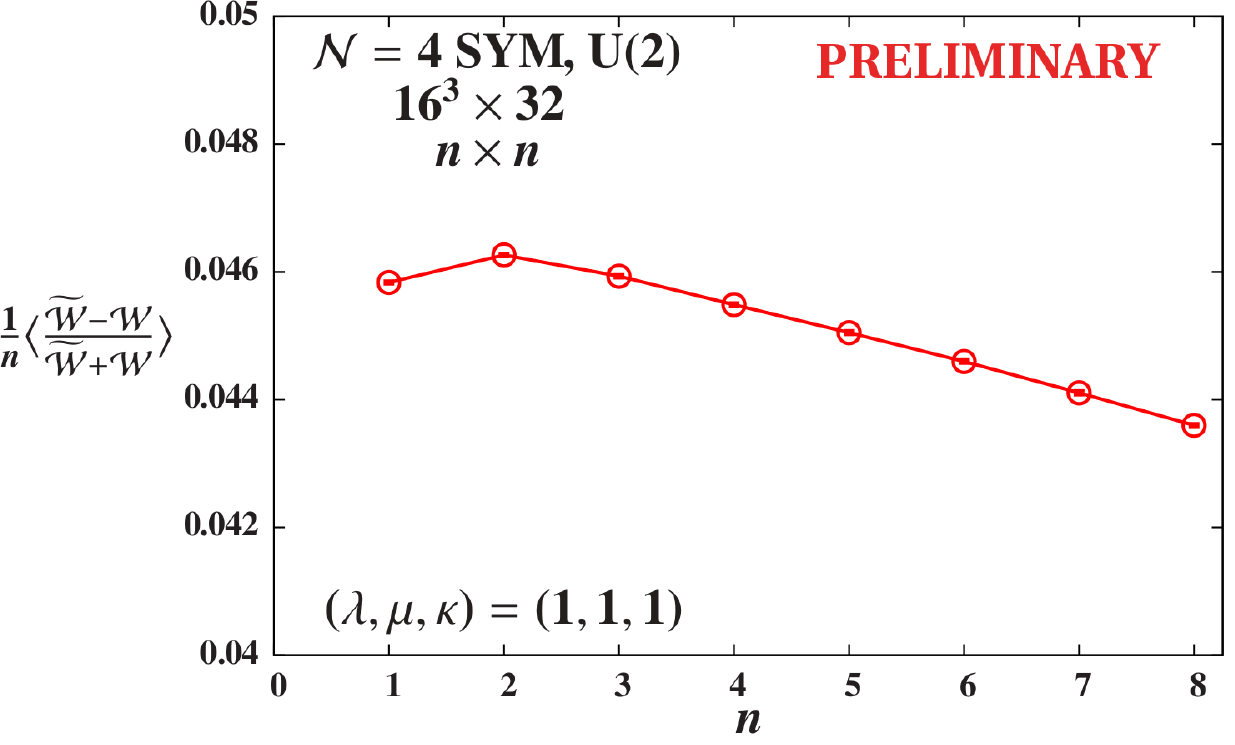}\hfill \includegraphics[width=0.45\linewidth]{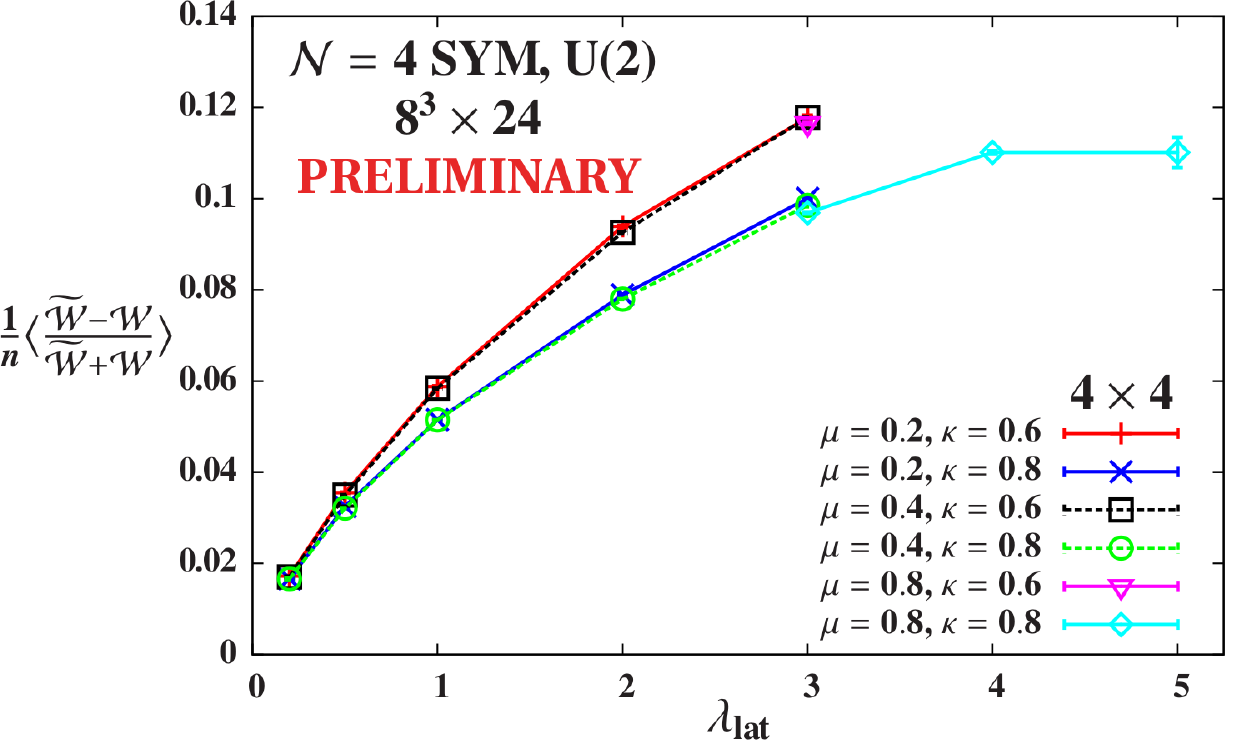}
  \caption{\label{fig:rsymm} Violations of $R_a$ symmetry, with lines connecting points to guide the eye.  {\bf Left:} For $n\X n$ loops on $16^3\X 32$ lattices with $(\lalat, \mu, \ka) = (1, 1, 1)$, the violations decrease (slightly) at larger distance scales, as expected.  {\bf Right:} For $4\X 4$ loops on $8^3\X 24$ lattices with various $(\mu, \ka)$, we find 5--10\% violations that vanish in the $\lalat \to 0$ limit as desired.}
\end{figure}

We present some results for these $R_a$ symmetry violations in \fig{fig:rsymm}.
A complication we encountered is that inverting the non-unitary links changes their gauge-invariant magnitudes,
\begin{equation}
  \Tr{\cUb_a^{-1} \cU_a^{-1}} > \Tr{\cU_a \cUb_a}.
\end{equation}
This causes the relative differences to be unreasonably sensitive to the number ($2n$) of inverted links in the $m\X n$ modified Wilson loops $\cWtw$.
The question of how best to handle this artifact is not yet settled.
We suspect that projecting \cW and \cWtw from U($N$) to SU($N$), by dividing out the determinant of each loop, will prove beneficial.
Such a determinant-divided analysis is not yet complete, so for the time being we empirically note that simply normalizing the relative differences by $\frac{1}{2n}$ stabilizes the results, leaving the expected decrease in violations as we approach the $1 / L \to 0$ continuum limit.
This is shown in the left panel of \fig{fig:rsymm}, where we can see that the resulting violations based on square $n\X n$ loops decrease only gradually on larger distance scales, falling by just $\sim$5\% as $n$ increases from 2 to 8 on $16^3\X32$ lattices.
This $\frac{1}{2n}$ normalization is not well motivated, and is likely to be replaced by a more robust prescription in the future.

The right panel of \fig{fig:rsymm} considers $R_a$ symmetry breaking based on $4\X 4$ loops for our main $8^3\X 24$ U(2) lattice ensembles, as functions of the 't~Hooft coupling.
These systems exhibit \mbox{5--10\%} violations that vanish in the $\lalat \to 0$ limit, as desired.
There is no visible sensitivity to $\mu$, but larger values of the U(1)-suppressing \ka parameter noticeably reduce $R_a$ symmetry breaking, which motivates the determinant projection to SU($N$) discussed above.
These results are similar to those based on the $1\X 1$ plaquette that we presented in \refcite{Catterall:2014vka}.

\begin{figure}[bp]
  \centering
  \includegraphics[height=\figheight]{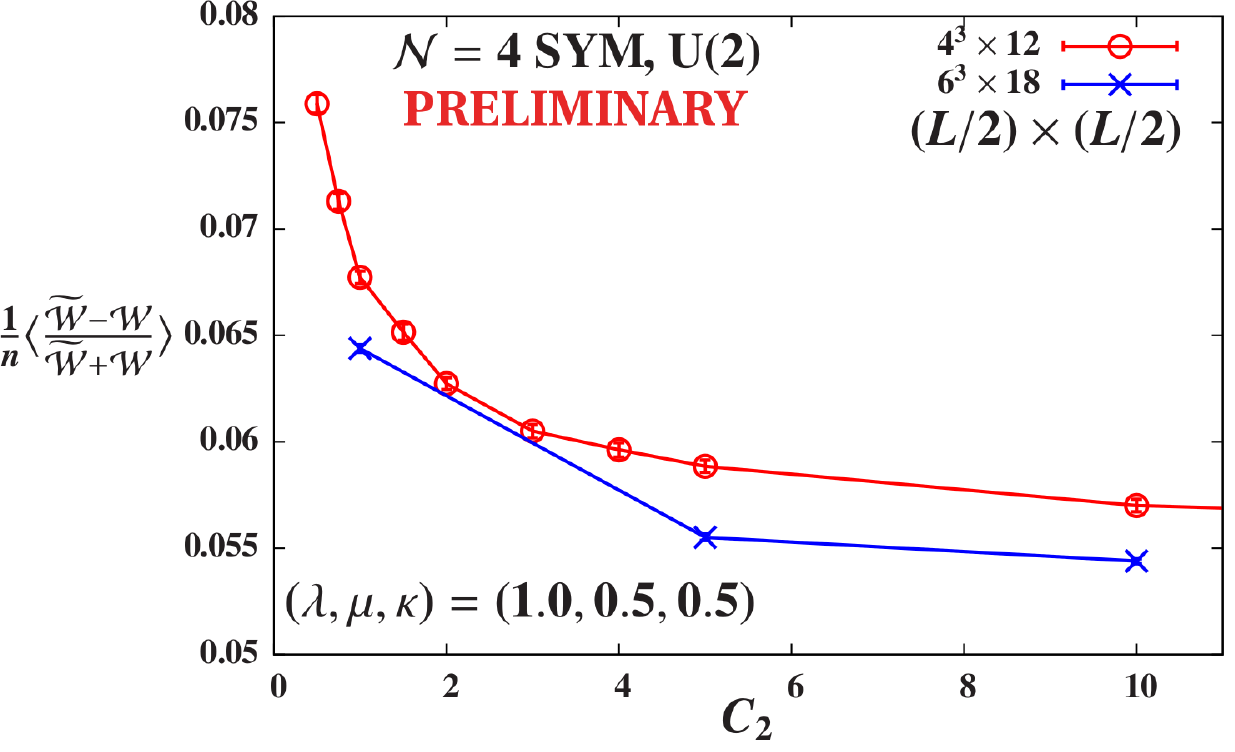}
  \caption{\label{fig:rsymm_c2} Study of tuning $c_2$ in \protect\eq{eq:c2}.  We consider $R_a$ symmetry breaking in $2\X2$ loops on $4^3\X 12$ lattices and $3\X 3$ loops on $6^3\X 18$ lattices, all with $(\lalat, \mu, \ka) = (1, 0.5, 0.5)$.  We observe gradual improvement as $c_2$ increases above its classical value $c_2 = 1$, while violations increase steeply for smaller $c_2 < 1$.  Lines connect points with the same volume to guide the eye.}
\end{figure}

Given this observable sensitive to the restoration of $\cQ_a$ and $\cQ_{ab}$, we can revisit the fine-tuning discussed in \secref{sec:Seff}.
Our goal is to choose the parameter $c_2$ in \eq{eq:c2} in order to minimize $R_a$ symmetry breaking and thereby approach the continuum theory as rapidly and smoothly as possible.
\fig{fig:rsymm_c2} presents an initial study of this tuning on small $4^3\X 12$ and $6^3\X 18$ lattices with fixed $(\lalat, \mu, \ka) = (1, 0.5, 0.5)$.
As desired, violations decrease for larger $3\X 3$ Wilson loops on the larger volume, though again the approach to the $1 / L \to 0$ continuum limit is gradual.
In addition we see that values of $c_2$ larger than the classical $c_2 = 1$ lead to $\sim$10\% smaller violations of $R_a$ symmetry, though on these volumes there is no strongly preferred value; the results flatten out as $c_2$ increases.
For $c_2 < 1$, in contrast, $R_a$ symmetry breaking increases steeply.
Conveniently, we also find that \cQ breaking behaves similarly (not shown), improving gradually for larger $c_2$. % TODO: Could add another s_B plot if there is extra space
This is important because restoration of the full symmetries of $\cN = 4$ SYM requires both \cQ and $R_a$.
For the time being we continue to use the classical $c_2 = 1$ in our larger-volume studies.
\fig{fig:rsymm_c2} suggests that larger $c_2$ would only result in modest improvement of $R_a$ symmetry breaking.

\begin{figure}[bp]
  \includegraphics[width=0.45\linewidth]{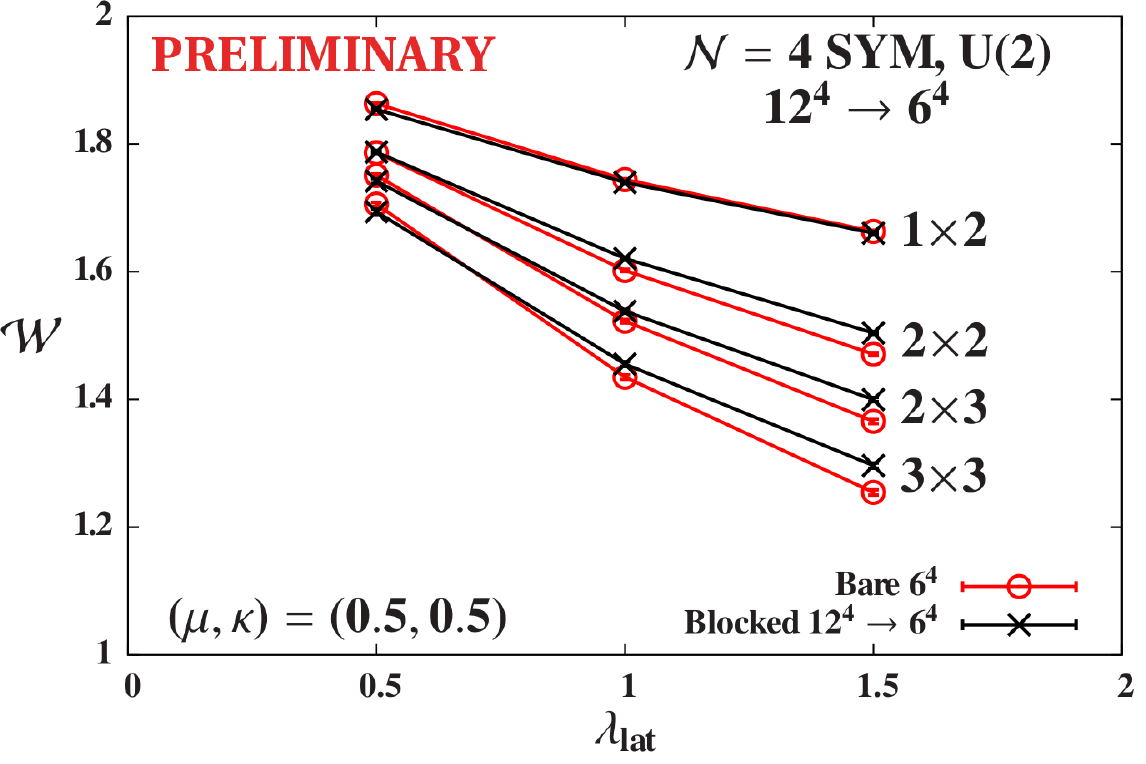}\hfill \includegraphics[width=0.49\linewidth]{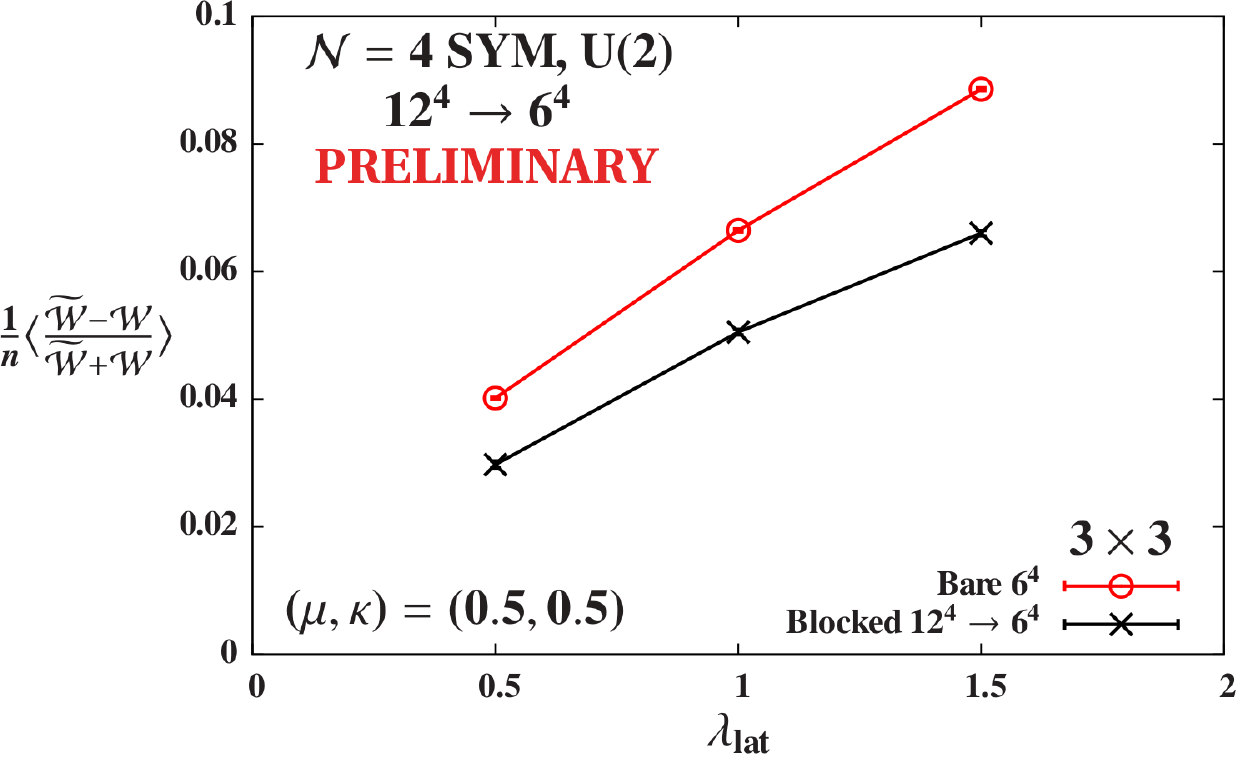}
  \caption{\label{fig:rsymm_MCRG} Wilson loops (left) and violations of $R_a$ symmetry (right) after one step of $12^4 \to 6^4$ MCRG blocking.  Lines connect points for loops of the same size to guide the eye.  The approximate agreement between the blocked Wilson loops and ``bare'' measurements on independent $6^4$ ensembles with the same couplings indicates that those couplings don't flow significantly under RG blocking.  The reduced $R_a$ symmetry breaking is consistent with the blocked measurements probing larger distance scales closer to the $1 / L \to 0$ continuum limit.}
\end{figure}

In \fig{fig:rsymm_MCRG} we present an initial exploration of MCRG blocking based on \eq{eq:block}.
We block $12^4$ lattices down to $6^4$ and measure observables on the blocked lattices as well as on independent $6^4$ ensembles with fixed $0.5 \leq \lalat \leq 1.5$ and $(\mu, \ka) = (0.5, 0.5)$.
As in \refcite{Catterall:2014mha}, we fix $\xi^4 \approx 1.2$ by requiring that the blocked plaquette matches the ``bare'' value on the $6^4$ ensembles. % TODO: Could add this plot if there is extra space
The left panel of \fig{fig:rsymm_MCRG} shows that all larger blocked Wilson loops then approximately match the corresponding bare loops, extending the results in \refcite{Catterall:2014mha}.
This evidence that the couplings don't flow noticeably under RG blocking is consistent with our observation of a Coulombic static potential in \secref{sec:potential}.
However, the blocked $R_a$ symmetry violations in the right panel of \fig{fig:rsymm_MCRG} end up significantly smaller than the bare $6^4$ results, by $\sim$25\% for these $3\X 3$ loops.
This is the behavior we expect as we move towards the $1 / L \to 0$ continuum conformal theory.

At present we have investigated the effects of the $R_a$ transformation only for the gauge links, through Eqs.~\ref{eq:Rtrans} and \ref{eq:Rsymm}.
Additional tests of the discrete R symmetries involving fermions should also be carried out in the future, to ensure that symmetry restoration is consistent across all sectors of the theory.
Unfortunately, the rescalings that we applied to the fermion fields in \eq{eq:rescale} interfere with such analyses, and further exploration is required to determine the best approach.
% ------------------------------------------------------------------

% ------------------------------------------------------------------
\section{\label{sec:pfaff}Measuring the complex Pfaffian to assess a potential sign problem} % Draft complete
A worrisome aspect of our numerical calculations is that they must be phase-quenched to take advantage of efficient importance sampling algorithms~\cite{Schaich:2014pda}.
Gaussian integration over the fermion fields $(\eta, \psi_a, \chi_{ab})$ in the path integral produces the Pfaffian of the fermion operator \cD defined by \eq{eq:Sexact}.
This Pfaffian is not manifestly real for any given gauge configuration, $\pf \cD = |\pf \cD|e^{i\al}$, and we retain only its magnitude in the path integral, omitting its phase $e^{i\al}$.
So long as $\vev{e^{i\al}}$ is nonzero, bona fide operator expectation values may be determined through phase reweighting.
If \al fluctuates so much that $\vev{e^{i\al}}$ is consistent with zero, then simulations would suffer from a sign problem and numerical results could not be trusted.

\begin{figure}[bp]
  \centering
  \includegraphics[height=\figheight]{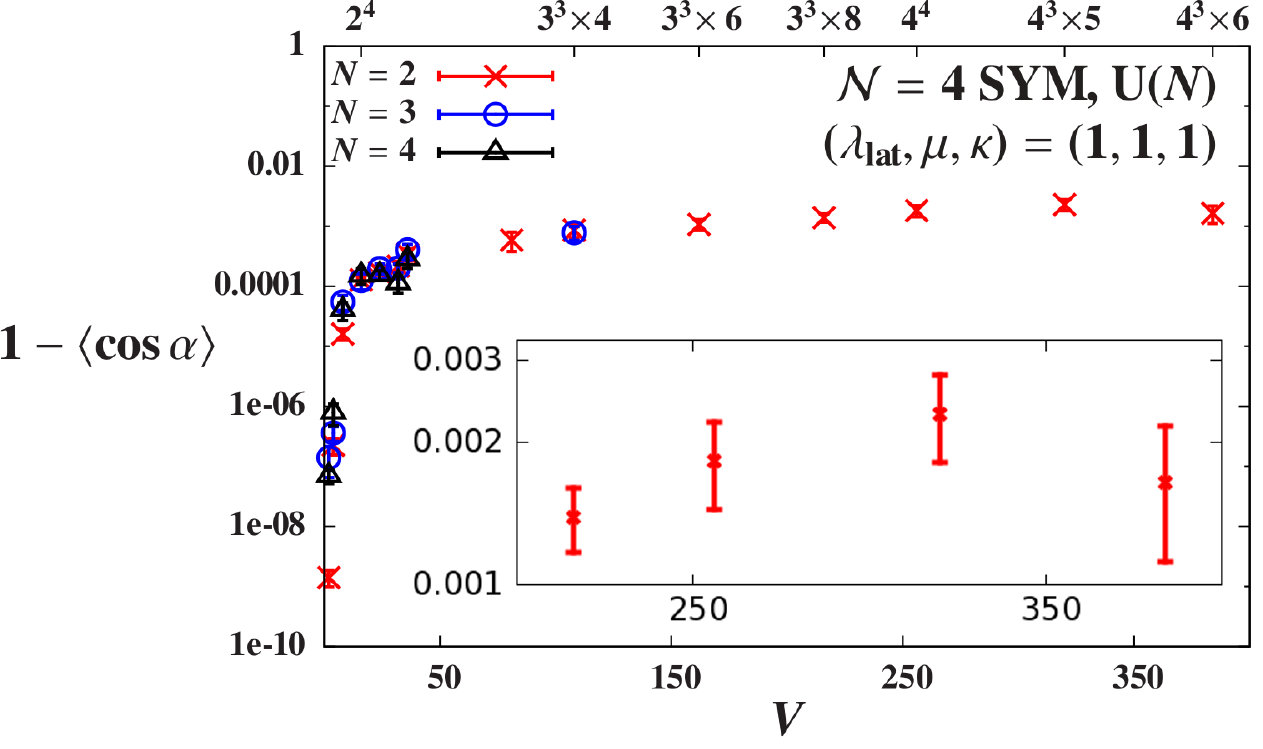}
  \caption{\label{fig:pfaffian} Semi-log plot of $1 - \vev{\cos\al}$ vs.\ lattice volume, where \al is the phase of the pfaffian.  All points are for $(\lalat, \mu, \ka) = (1, 1, 1)$, for gauge groups U(2), U(3) and U(4).  The inset zooms in on the four largest-volume U(2) results with $V \geq 3^3\X 8$, where the small phase does not grow with volume.}
\end{figure}

In \refcite{Catterall:2014vka} we initiated an ongoing study of the complex Pfaffian of lattice $\cN = 4$ SYM.
Focusing on gauge group U(2) we found that $e^{i\al}$ was approximately real and positive on every investigated system, close enough to unity that phase reweighting is not necessary to obtain reliable expectation values.
Fluctuations in the phase did not grow with the system size on the largest accessible volumes up to $4^3\X6$, and on $2^3\X4$ lattices $\vev{\cos\al}$ showed little dependence on the number of colors $N = 2$, 3 or 4.
These results are included in \fig{fig:pfaffian}, which also considers additional U(3) and U(4) lattice volumes, finding similar behavior for all gauge groups.
All points in this figure are for $(\lalat, \mu, \ka) = (1, 1, 1)$.
In initial investigations we have observed even better behavior for smaller \lalat and no significant dependence on $\ka$, so long as \ka is large enough to suppress the U(1) confinement transition discussed in \secref{sec:stab}.

Evaluating the Pfaffian is a notoriously hard problem, and these results are only obtainable thanks to new parallel software described in detail by \refcite{Schaich:2014pda}.
Despite this new code, we remain limited to lattice volumes around $4^3\X6$ (and smaller for larger $N$), where several days are already required for each measurement.
However, now that we have designed and implemented the RG blocking transformation in \eq{eq:block}, it will be interesting to measure the Pfaffian on lattices blocked down to a small number of sites.
In addition to searching for information about the phase on our larger-volume lattice ensembles, by comparing systems after different numbers of RG blocking steps we may gain insight into the behavior of the Pfaffian in the continuum limit.
% ------------------------------------------------------------------

% ------------------------------------------------------------------
\section{\label{sec:potential}Static potentials and Coulomb coefficients} % Draft complete
In previous sections we focused on validating our lattice system and phase-quenched calculations, to ensure that we simulate $\cN = 4$ SYM to a good approximation and can recover the appropriate theory in the continuum limit.
This work is necessary to establish that numerical results from lattice $\cN = 4$ SYM can be trusted.
Let us now consider a more physical quantity, the static potential.
In \refcite{Catterall:2014vka} we presented results for the U(2) potential, finding Coulombic behavior $V(r) = A - C / r$ at both weak and strong 't~Hooft coupling, with the Coulomb coefficients $C$ in good agreement with leading-order perturbation theory.
Here we supplement these results with a first look at the U(3) static potential.

We extract the static potential from Wilson loops $W(\vec r, t)$ measured on $8^3\X 24$ lattices.
In \secref{sec:susies} we considered $m\X n$ Wilson loops oriented along the principal axes of the lattice.
For the static potential we want to be more general and consider all possible spatial separations $\vec r$, which we do by gauge fixing to Coulomb gauge and computing
\begin{equation}
  \label{eq:Wcorr}
  W(\vec r, t) = \Tr{P(\vec x, t, t_0) P^{\dag}(\vec x + \vec r, t, t_0)}.
\end{equation}
Here $P(\vec x, t, t_0)$ is a product of temporal links $\cU_t$ at spatial location $\vec x$, extending from timeslice $t_0$ to timeslice $t_0 + t$.
We extract $V(r)$ by fitting $W(r, t) = w \exp(-V(r) t)$, combining all $\vec r$ with the same magnitude $r \equiv |\vec r|$ and comparing fit ranges $t_{min} \leq t < N_t / 2 = 12$ to find the $t_{min} = 5$--6 above which the results form stable plateaus.

\begin{figure}[bp]
  \includegraphics[width=0.45\linewidth]{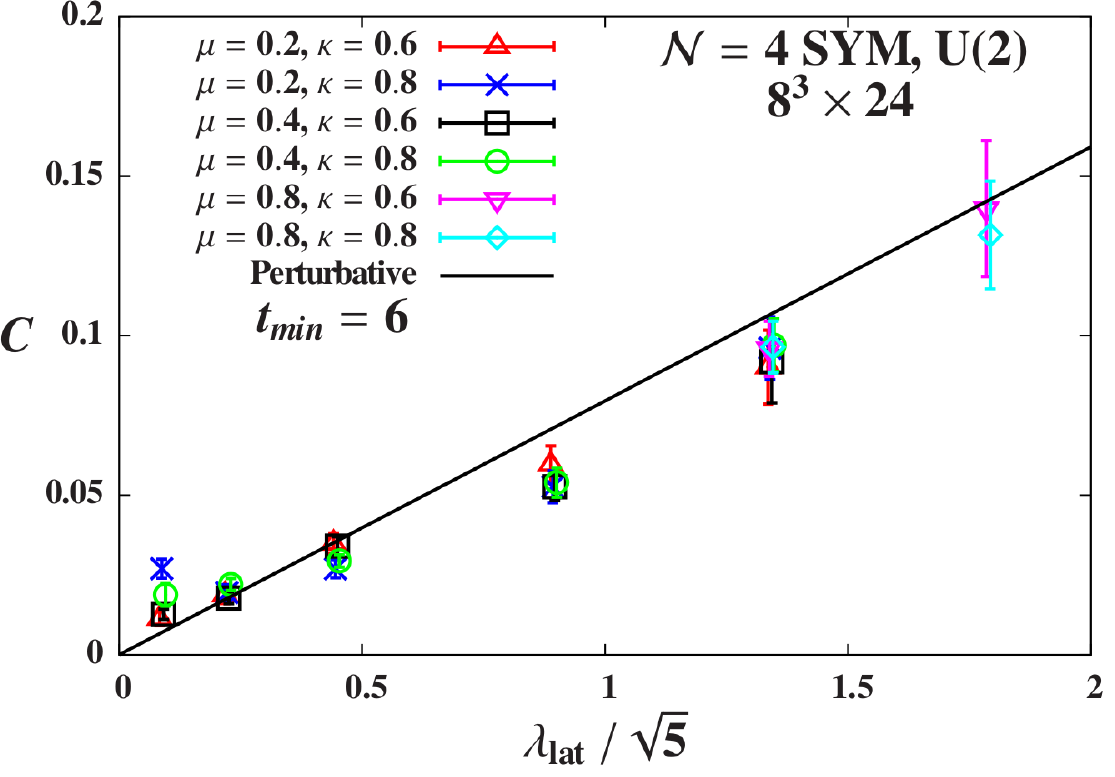}\hfill \includegraphics[width=0.45\linewidth]{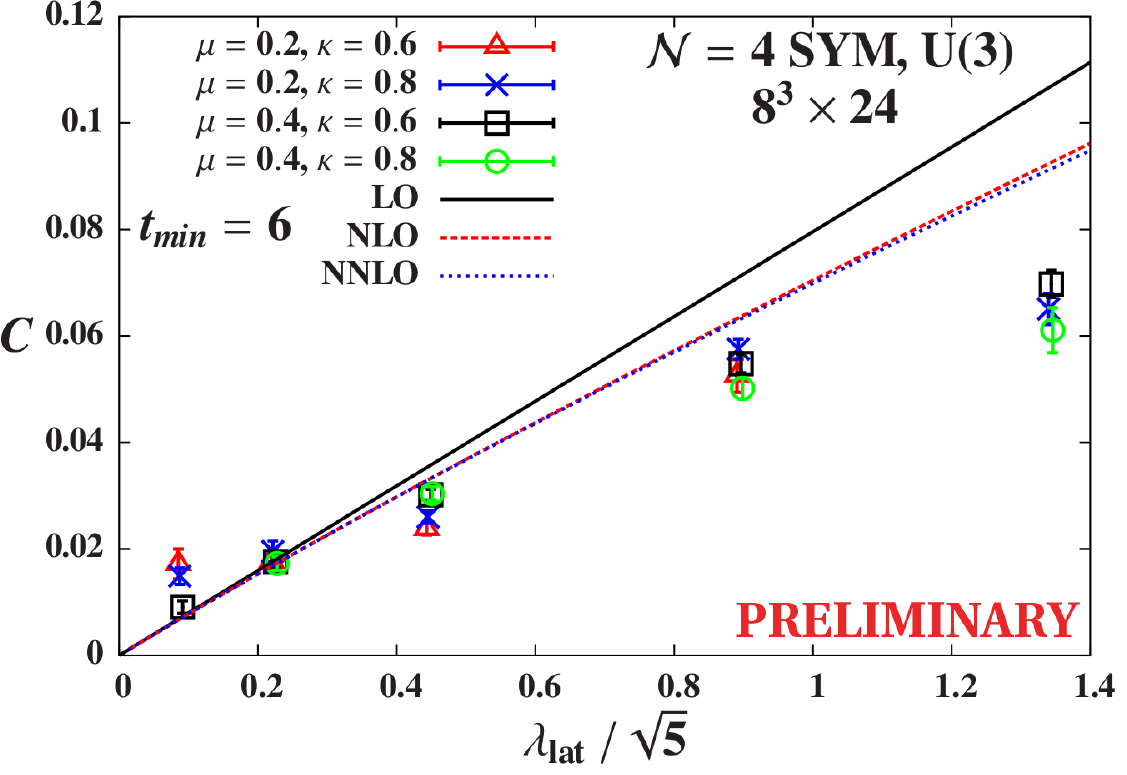}
  \caption{\label{fig:C} Static potential Coulomb coefficients $C$ for the usual Wilson loops from $8^3\X 24$ lattices with gauge groups U(2) (left, from \protect\refcite{Catterall:2014vka}) and U(3) (right).  The $N = 2$ results are consistent with the leading-order perturbative prediction.  The $N = 3$ results fall below even the NNLO perturbative prediction from Refs.~\protect\cite{Pineda:2007kz, Stahlhofen:2012zx, Prausa:2013qva}.}
\end{figure}

To directly compare our data and results with continuum expectations, we must take care to account for the physical structure of the $A_4^*$ lattice.
As discussed in \refcite{Catterall:2014vka} we use a convenient representation of the $A_4^*$ lattice that adds the fifth link $\muhat_4 = (-1, -1, -1, -1)$ to the usual four hypercubic basis vectors.
This allows us to use familiar integer displacement vectors in the code, which must be converted to physical distances on the true $A_4^*$ lattice.
Doing so, we find that displacements corresponding to any $r > 2.6$ can wrap around the $L = 8$ spatial lattice, making them unusable in the gauge-fixed correlator \eq{eq:Wcorr}.
Limited investigations of $L = 12$ and 16 suggest that our signals disappear into the background noise for $r \gtrsim 3$ in any case, so that we don't lose much due to the restriction $\sqrt{3 / 4} \leq r \leq 2.6$ on our $8^3\X 24$ lattices.

An additional consequence of the $A_4^*$ lattice structure is the normalization factor relating the continuum and lattice couplings, $\la = \lalat / \sqrt 5$.
This is simply the Jacobian of the transformation between the lattice and continuum space-time coordinates, which is non-trivial because the $A_4^*$ links are not orthogonal.
We provide a more intuitive derivation of this rescaling in \refcite{Catterall:2014vka}.

Fitting $V(r)$ to the Coulomb form $V(r) = A - C / r$, using all accessible $\sqrt{3 / 4} \leq r \leq 2.6$, produces the Coulomb coefficient results shown in \fig{fig:C}. % TODO: Could add V(r) plots if there is extra space
For each gauge group U(2) and U(3) we consider several values of $(\mu, \ka)$, observing no significant dependence on these parameters.
We plot the Coulomb coefficients as functions of the continuum 't~Hooft coupling $\la = \lalat / \sqrt 5$, and compare our numerical results with perturbative predictions.
For $N = 2$ in the left panel of \fig{fig:C} we find values of $C$ in good agreement with leading-order (LO) perturbation theory,
\begin{equation}
  \label{eq:CLO}
  C_{LO} = \frac{\lalat / \sqrt 5}{4\pi} = \frac{\la}{4\pi}.
\end{equation}
For $N = 3$ in the right panel, however, our results begin deviating significantly from the LO line at stronger couplings.
Although the next-to-leading-order and next-to-next-to-leading-order corrections from Refs.~\cite{Pineda:2007kz, Stahlhofen:2012zx, Prausa:2013qva} move the perturbative prediction closer to our results, a clear difference remains.
These NLO and NNLO contributions are
\begin{align}
  \frac{C_{NLO}}{C_{LO}}  & = \frac{\la}{2\pi^2} \left(\log\left[\frac{\la}{2\pi}\right] + \ga_E - 1\right)                                                                                                                                                   \label{eq:Cpert} \\
  \frac{C_{NNLO}}{C_{LO}} & = \frac{\la^2}{8\pi^4} \left\{\left(\log\left[\frac{\la}{2\pi}\right] + \ga_E\right)^2 + \left(1 + \frac{\pi^2}{3}\right)\left(\log\left[\frac{\la}{2\pi}\right] + \ga_E\right) - \frac{\pi^2}{12}  - \frac{7}{2} + \frac{9}{4}\zeta(3)\right\}  \nonumber
\end{align}
where $\ga_E$ is Euler's constant, $\zeta(3) \approx 1.202$ is Ap\'ery's constant, and we have set $C_A \al = 2C_F \al = \frac{\la}{4\pi}$ for gauge group U($N$).

\begin{figure}[bp]
  \includegraphics[width=0.45\linewidth]{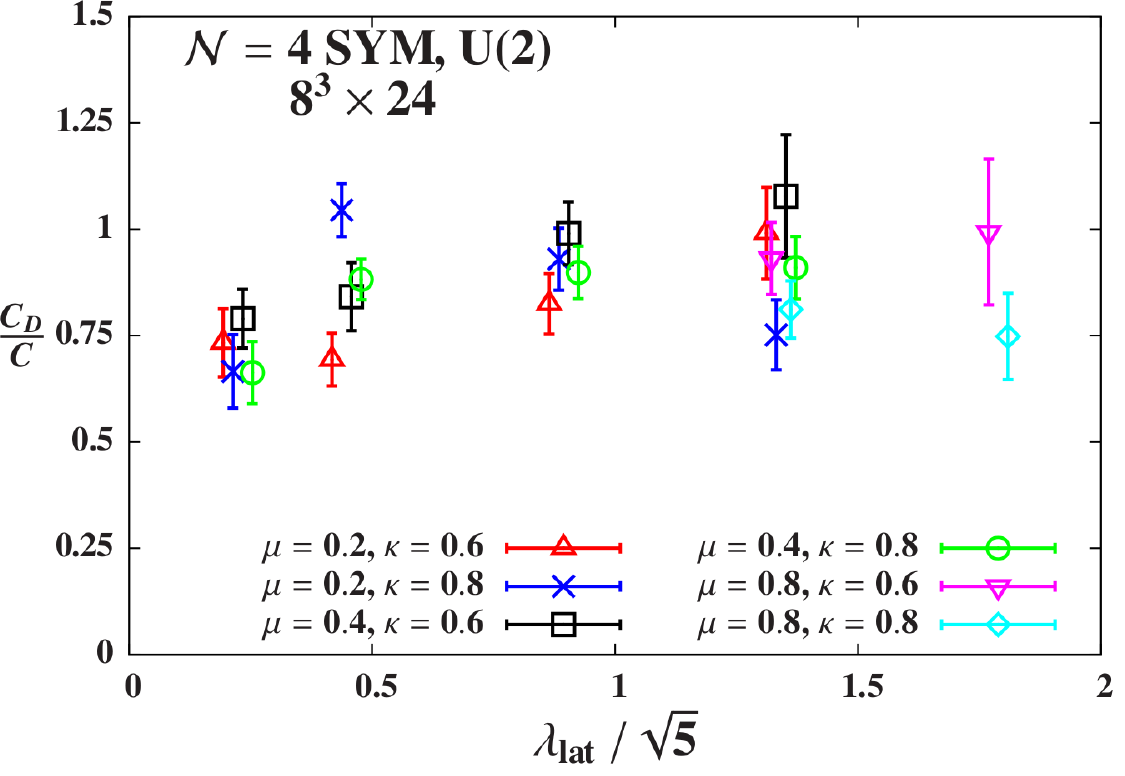}\hfill \includegraphics[width=0.45\linewidth]{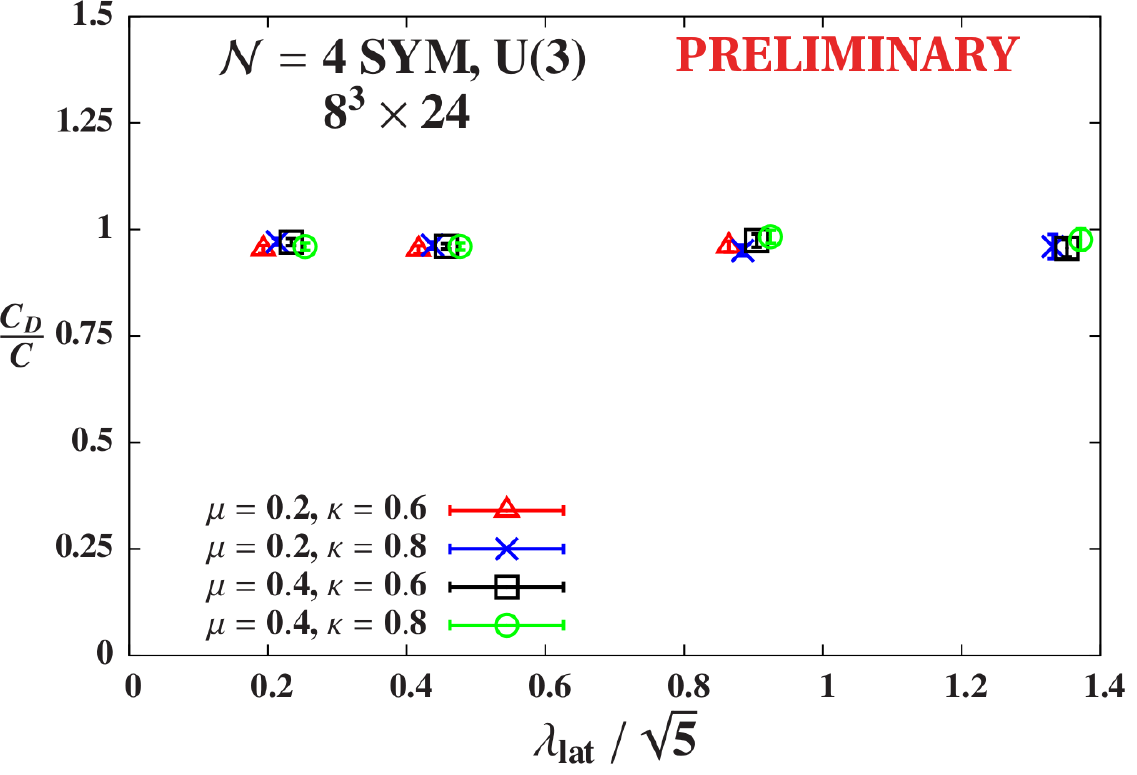}
  \caption{\label{fig:Cdet} Ratios $C_D / C$ of static potential Coulomb coefficients for the determinant-divided Wilson loops relative to those for the usual loops, from $8^3\X 24$ lattices with gauge groups U(2) (left, from \protect\refcite{Catterall:2014vka}) and U(3) (right).  Both plots show the same vertical range to illustrate that the new $N = 3$ results are much less noisy than $N = 2$, though the former are consistently larger than the expected ratio $1 - \frac{1}{N^2} = \frac{8}{9}$.}
\end{figure}

While it is tempting to interpret the right panel of \fig{fig:C} as a suggestion that we are seeing $C$ depart from perturbation theory to approach the famous large-$N$ prediction $C \propto \sqrt{\la}$ at strong coupling~\cite{Rey:1998ik, Maldacena:1998im}, we suspect that both $N = 3$ and $\la \leq 3 / \sqrt 5$ are still too small for this to happen.
We can see that the NLO and (especially) NNLO corrections are still small compared to $C_{LO}$, suggesting that our results should remain within the perturbative regime.
Although the contrast between $C$ for U(2) and U(3) in the two plots of \fig{fig:C} is striking, its cause is not yet clear.
It will be very interesting to add U(4) results to this picture, to clarify any systematic trends.
This work is underway, but is challenging due to the steeply increasing computation costs, which grow $\propto N^5$~\cite{Schaich:2014pda}.

Another interesting difference between the U(2) and U(3) static potentials is visible when we divide out the determinant of the Wilson loops from which $V(r)$ is extracted.
As discussed in \secref{sec:susies}, this procedure should approximately project the observables to SU($N$), removing the U(1) sector that decouples in the continuum limit.
We expect that the Coulomb coefficients $C_D$ resulting from determinant-divided Wilson loops will be reduced by a factor $(N^2 - 1) / N^2$ compared to the usual $C$ in \fig{fig:C}.
(This holds in perturbation theory through NLO, and the departure from this simple scaling at NNLO~\cite{Prausa:2013qva} appears negligible.)

In \fig{fig:Cdet} we test our expectation by plotting the ratios $C_D / C$, again as functions of the continuum $\la = \lalat / \sqrt 5$ for several values of $(\mu, \ka)$.
As before, there is no clear dependence on $(\mu, \ka)$, and the ratios are also insensitive to the coupling.
Our U(2) results are fairly noisy, but appear consistent with the expected ratio of $3 / 4$.
In contrast, our results for $N = 3$ are much more stable, which we emphasize by showing the same vertical range in both plots of \fig{fig:Cdet}.
Numerically we find $C_D / C \approx 0.96$ with uncertainties small enough to clearly differ from the expected ratio of $8 / 9$.
Again, ongoing investigations of U(4) will be important to clarify the interpretation of these initial U(3) results.

\begin{figure}[bp]
  \includegraphics[width=0.45\linewidth]{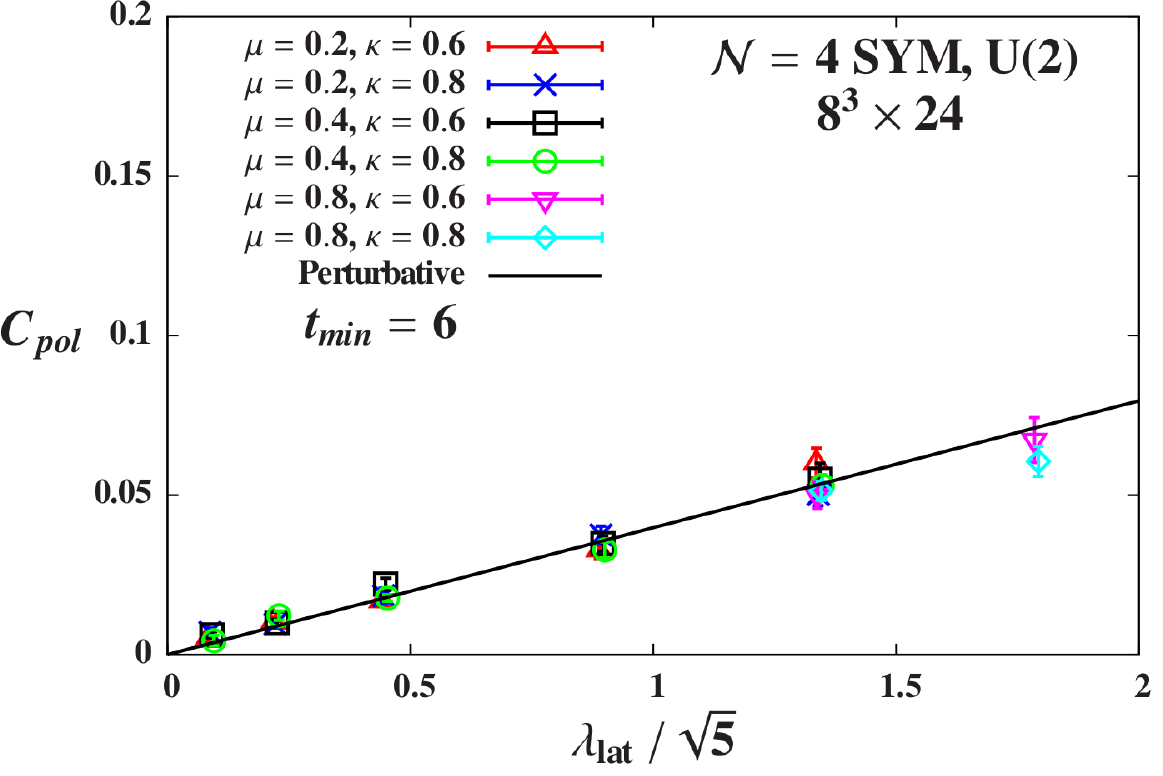}\hfill \includegraphics[width=0.45\linewidth]{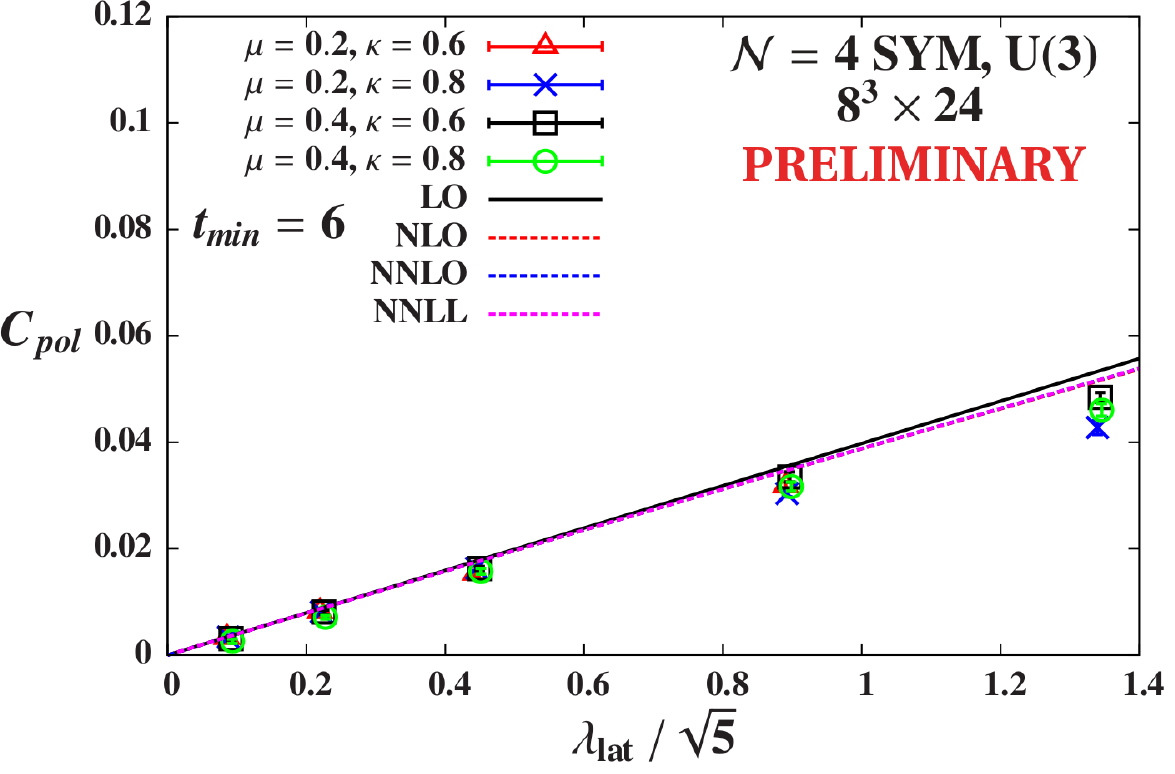}
  \caption{\label{fig:Cpol} Static potential Coulomb coefficients $C_{pol}$ for the polar-projected Wilson loops that decouple the scalar fields, from $8^3\X 24$ lattices with gauge groups U(2) (left, from \protect\refcite{Catterall:2014vka}) and U(3) (right).  The $N = 2$ results are consistent with the leading-order perturbative prediction, and the $N = 3$ results also remain close to perturbation theory.}
\end{figure}

Finally, we can also consider the $\cN = 4$ static potential with the contribution of the scalar fields removed~\cite{Pineda:2007kz, Prausa:2013qva}.
We implement this by building Wilson loops from unitary matrices $u$ defined through the polar decomposition of each link variable, $\cU = u H$ where $H$ is Hermitian and positive definite.
Decoupling the scalar fields simplifies the perturbative predictions for the Coulomb coefficient in Refs.~\cite{Pineda:2007kz, Prausa:2013qva}, with Eqs.~\ref{eq:CLO} and \ref{eq:Cpert} becoming
\begin{equation}
  \begin{split}
  C_{pol} & = \frac{\la}{8\pi}\left\{1 - \frac{\la}{4\pi^2} + A\left(\frac{\la}{4\pi^2}\right)^2 + B\left(\frac{\la}{4\pi^2}\right)^3\log\left[\frac{\la}{4\pi}\right] + \cO\left(\la^3\right)\right\} \\
  & \hspace{2 cm} A = \frac{5}{4} + \frac{\pi^2}{4} - \frac{\pi^4}{64} \hspace{2 cm} B = \frac{1}{12\pi}.
  \end{split}
\end{equation}
At leading order the polar-projected $C_{pol}$ is just half the usual Coulomb coefficient, $C_{pol} / C = 1 / 2$.

In \fig{fig:Cpol} we plot $C_{pol}$ vs.\ $\lalat / \sqrt 5$ for $N = 2$ and 3, using the same axes and considering the same values of $(\mu, \ka)$ as in \fig{fig:C}.
The U(2) results for $C_{pol}$ in the left panel of \fig{fig:Cpol} are in good agreement with leading-order perturbation theory, and appear less noisy than the usual Coulomb coefficients in \fig{fig:C}.
The U(3) results in the right panel of \fig{fig:Cpol} are significantly closer to leading-order perturbation theory than the corresponding $C$ in \fig{fig:C}.
At the same time, the higher-order perturbative corrections have barely visible effects in this range of $\lalat$.
The $N = 3$ $C_{pol}$ may still begin to drop below the perturbative predictions at the strongest coupling we have investigated so far, but this deviation is not significant and may be cured by extrapolating $(\mu, \ka) \to (0, 0)$.
% ------------------------------------------------------------------

% ------------------------------------------------------------------
\section{Recapitulation of status, prospects and next steps} % Draft complete
Our investigations of lattice $\cN = 4$ SYM are making progress addressing a number of interesting questions, regarding both the validity of our lattice calculations as well as the resulting predictions that may be compared to continuum perturbation theory and gauge--gravity duality.
In this proceedings, we presented several new developments that go beyond the results appearing in Refs.~\cite{Catterall:2014vka, Catterall:2014mha}.
These include an initial study of fine-tuning to recover the full symmetries of $\cN = 4$ SYM in the continuum limit, which makes use of the recent derivation that only a single parameter needs to be tuned so long as the moduli space is not lifted by nonperturbative effects.
We also extended our measurements of the complex Pfaffian, considering more lattice volumes for larger $N = 3$ and 4, which also indicate that the lattice theory does not suffer from a sign problem.
These results provide additional evidence that our phase-quenched lattice calculations simulate $\cN = 4$ SYM to a good approximation, and can recover the appropriate theory in the continuum limit.

We then focused on the static potential, extending our previous study of gauge group U(2) with initial results for $N = 3$.
In both cases, we find Coulombic behavior at both weak and strong coupling, and compare the Coulomb coefficients to perturbation theory.
Our initial U(3) results appear significantly less noisy than those for $N = 2$, but fall away from perturbative predictions at the strongest couplings we have investigated.
When we consider Wilson loops approximately projected to SU(3), we find Coulomb coefficients $C_D$ closer to the usual U(3) $C$ than we expected from naive $(N^2 - 1) / N^2$ scaling.
It is not yet clear how we should interpret the current contrast between $N = 2$ and 3.
We are working on extending these investigations to $N = 4$, which we hope will clarify any systematic trends.

Additional future work will include further studies of the Pfaffian phase, with the aim of improving our qualitative understanding of why we see no sign problem for lattice $\cN = 4$ SYM.
Although it is extremely expensive to measure the Pfaffian on larger lattice volumes or for larger values of $N$, we may be able to gain insight by applying our new RG blocking scheme to obtain blocked systems with fewer degrees of freedom, which are more practical to analyze.
MCRG analyses may also prove useful to predict the anomalous dimensions of single-trace operators like the Konishi, a particularly exciting goal.
Such predictions would be complementary to those obtainable in perturbation theory~\cite{Beem:2013hha} or via the conformal bootstrap program~\cite{Beem:2013qxa}.
% ------------------------------------------------------------------

% ------------------------------------------------------------------
\vspace{12 pt}
\noindent {\sc Acknowledgments:}~We thank Maximilian Stahlhofen for assistance with higher-order perturbation theory for the static potential.
This work was supported in part by the U.S.~Department of Energy (DOE), Office of Science, Office of High Energy Physics, under Award Numbers DE-SC0008669 (DS), DE-SC0009998 (SC, DS), DE-SC0010005 (TD) and DE-FG02-08ER41575 (JG).
Numerical calculations were carried out on the HEP-TH cluster at the University of Colorado and on the DOE-funded USQCD facilities at Fermilab.
% ------------------------------------------------------------------

% ------------------------------------------------------------------
\bibliographystyle{utphys}
\bibliography{lattice2014}

\providecommand{\href}[2]{#2}\begingroup\raggedright\begin{thebibliography}{10}

\bibitem{Catterall:2009it}
S.~Catterall, D.~B. Kaplan and M.~Unsal, ``{Exact lattice supersymmetry}'',
  \href{http://dx.doi.org/10.1016/j.physrep.2009.09.001}{{\em Phys. Rept.} {\bf
  484} (2009) 71--130} [\href{http://arxiv.org/abs/0903.4881}{{\tt
  arXiv:0903.4881}}].

\bibitem{Ishii:2008ib}
T.~Ishii, G.~Ishiki, S.~Shimasaki and A.~Tsuchiya, ``{$\mathcal N = 4$ Super
  Yang--Mills from the Plane Wave Matrix Model}'',
  \href{http://dx.doi.org/10.1103/PhysRevD.78.106001}{{\em Phys. Rev.} {\bf
  D78} (2008) 106001} [\href{http://arxiv.org/abs/0807.2352}{{\tt
  arXiv:0807.2352}}].

\bibitem{Ishiki:2008te}
G.~Ishiki, S.-W. Kim, J.~Nishimura and A.~Tsuchiya, ``{Deconfinement phase
  transition in $\mathcal N = 4$ super Yang--Mills theory on $R\times S^3$ from
  supersymmetric matrix quantum mechanics}'',
  \href{http://dx.doi.org/10.1103/PhysRevLett.102.111601}{{\em Phys. Rev.
  Lett.} {\bf 102} (2009) 111601} [\href{http://arxiv.org/abs/0810.2884}{{\tt
  arXiv:0810.2884}}].

\bibitem{Ishiki:2009sg}
G.~Ishiki, S.-W. Kim, J.~Nishimura and A.~Tsuchiya, ``{Testing a novel large-N
  reduction for $\mathcal N = 4$ super Yang--Mills theory on $R\times S^3$}'',
  \href{http://dx.doi.org/10.1088/1126-6708/2009/09/029}{{\em JHEP} {\bf 0909}
  (2009) 029} [\href{http://arxiv.org/abs/0907.1488}{{\tt arXiv:0907.1488}}].

\bibitem{Hanada:2010kt}
M.~Hanada, S.~Matsuura and F.~Sugino, ``{Two-dimensional lattice for
  four-dimensional $\mathcal N = 4$ supersymmetric Yang--Mills}'',
  \href{http://dx.doi.org/10.1143/PTP.126.597}{{\em Prog. Theor. Phys.} {\bf
  126} (2011) 597--611} [\href{http://arxiv.org/abs/1004.5513}{{\tt
  arXiv:1004.5513}}].

\bibitem{Honda:2011qk}
M.~Honda, G.~Ishiki, J.~Nishimura and A.~Tsuchiya, ``{Testing the AdS/CFT
  correspondence by Monte Carlo calculation of BPS and non-BPS Wilson loops in
  4d $\mathcal N = 4$ super-Yang--Mills theory}'',
  \href{http://pos.sissa.it/archive/conferences/139/244/Lattice
  2011_244.pdf}{{\em PoS} {\bf Lattice 2011} (2011) 244}
  [\href{http://arxiv.org/abs/1112.4274}{{\tt arXiv:1112.4274}}].

\bibitem{Honda:2013nfa}
M.~Honda, G.~Ishiki, S.-W. Kim, J.~Nishimura and A.~Tsuchiya, ``{Direct test
  of the AdS/CFT correspondence by Monte Carlo studies of $\mathcal N = 4$
  super Yang--Mills theory}'',
  \href{http://dx.doi.org/10.1007/JHEP11(2013)200}{{\em JHEP} {\bf 1311} (2013)
  200} [\href{http://arxiv.org/abs/1308.3525}{{\tt arXiv:1308.3525}}].

\bibitem{Hanada:2013rga}
M.~Hanada, Y.~Hyakutake, G.~Ishiki and J.~Nishimura, ``{Holographic
  description of quantum black hole on a computer}'',
  \href{http://dx.doi.org/10.1126/science.1250122}{{\em Science} {\bf 344}
  (2014) 882--885} [\href{http://arxiv.org/abs/1311.5607}{{\tt
  arXiv:1311.5607}}].

\bibitem{Catterall:2011pd}
S.~Catterall, E.~Dzienkowski, J.~Giedt, A.~Joseph and R.~Wells,
  ``{Perturbative renormalization of lattice $\mathcal N = 4$ super Yang--Mills
  theory}'', \href{http://dx.doi.org/10.1007/JHEP04(2011)074}{{\em JHEP} {\bf
  1104} (2011) 074} [\href{http://arxiv.org/abs/1102.1725}{{\tt
  arXiv:1102.1725}}].

\bibitem{Catterall:2012yq}
S.~Catterall, P.~H. Damgaard, T.~Degrand, R.~Galvez and D.~Mehta, ``{Phase
  Structure of Lattice $\mathcal N = 4$ Super Yang--Mills}'',
  \href{http://dx.doi.org/10.1007/JHEP11(2012)072}{{\em JHEP} {\bf 1211} (2012)
  072} [\href{http://arxiv.org/abs/1209.5285}{{\tt arXiv:1209.5285}}].

\bibitem{Catterall:2013roa}
S.~Catterall, J.~Giedt and A.~Joseph, ``{Twisted supersymmetries in lattice
  $\mathcal N = 4$ super Yang--Mills theory}'',
  \href{http://dx.doi.org/10.1007/JHEP10(2013)166}{{\em JHEP} {\bf 1310} (2013)
  166} [\href{http://arxiv.org/abs/1306.3891}{{\tt arXiv:1306.3891}}].

\bibitem{Catterall:2014vka}
S.~Catterall, D.~Schaich, P.~H. Damgaard, T.~DeGrand and J.~Giedt,
  ``{$\mathcal N = 4$ supersymmetry on a space-time lattice}'',
  \href{http://dx.doi.org/10.1103/PhysRevD.90.065013}{{\em Phys. Rev.} {\bf
  D90} (2014) 065013} [\href{http://arxiv.org/abs/1405.0644}{{\tt
  arXiv:1405.0644}}].

\bibitem{Catterall:2014mha}
S.~Catterall and J.~Giedt, ``{Real space renormalization group for twisted
  lattice $\mathcal N = 4$ super Yang--Mills}'', {\em JHEP} (in press, 2014)
  [\href{http://arxiv.org/abs/1408.7067}{{\tt arXiv:1408.7067}}].

\bibitem{Schaich:2014pda}
D.~Schaich and T.~DeGrand, ``{Parallel software for lattice $\mathcal N = 4$
  supersymmetric Yang--Mills theory}'',
  \href{http://arxiv.org/abs/1410.6971}{{\tt arXiv:1410.6971}}.

\bibitem{Marcus:1995mq}
N.~Marcus, ``{The other topological twisting of $\mathcal N = 4$
  Yang--Mills}'', \href{http://dx.doi.org/10.1016/0550-3213(95)00389-A}{{\em
  Nucl. Phys.} {\bf B452} (1995) 331--345}
  [\href{http://arxiv.org/abs/hep-th/9506002}{{\tt hep-th/9506002}}].

\bibitem{Kapustin:2006pk}
A.~Kapustin and E.~Witten, ``{Electric-Magnetic Duality And The Geometric
  Langlands Program}'',
  \href{http://dx.doi.org/10.4310/CNTP.2007.v1.n1.a1}{{\em Commun. Num. Theor.
  Phys.} {\bf 1} (2007) 1--236}
  [\href{http://arxiv.org/abs/hep-th/0604151}{{\tt hep-th/0604151}}].

\bibitem{Unsal:2006qp}
M.~{\"U}nsal, ``{Twisted supersymmetric gauge theories and orbifold
  lattices}'', \href{http://dx.doi.org/10.1088/1126-6708/2006/10/089}{{\em
  JHEP} {\bf 0610} (2006) 089}
  [\href{http://arxiv.org/abs/hep-th/0603046}{{\tt hep-th/0603046}}].

\bibitem{Palumbo:1990kh}
F.~Palumbo, ``{Gauge Invariance on the Lattice With Noncompact Gauge Fields}'',
  \href{http://dx.doi.org/10.1016/0370-2693(90)90268-B}{{\em Phys. Lett.} {\bf
  B244} (1990) 55--57}.

\bibitem{Becchi:1992um}
C.~M. Becchi and F.~Palumbo, ``{Noncompact gauge theories on a lattice:
  Perturbative study of the scaling properties}'',
  \href{http://dx.doi.org/10.1016/0550-3213(92)90555-P}{{\em Nucl. Phys.} {\bf
  B388} (1992) 595--608}.

\bibitem{Palumbo:2001br}
F.~Palumbo and R.~Scimia, ``{Noncompact gauge fields on a lattice: SU($N$)
  theories}'', \href{http://dx.doi.org/10.1103/PhysRevD.65.074509}{{\em Phys.
  Rev.} {\bf D65} (2002) 074509}
  [\href{http://arxiv.org/abs/hep-lat/0105029}{{\tt hep-lat/0105029}}].

\bibitem{DeGrand:1980eq}
T.~A. DeGrand and D.~Toussaint, ``{Topological Excitations and Monte Carlo
  Simulation of Abelian Gauge Theory}'',
  \href{http://dx.doi.org/10.1103/PhysRevD.22.2478}{{\em Phys. Rev.} {\bf D22}
  (1980) 2478}.

\bibitem{Pineda:2007kz}
A.~Pineda, ``{Static potential in $\mathcal N = 4$ supersymmetric Yang-Mills at
  weak coupling}'', \href{http://dx.doi.org/10.1103/PhysRevD.77.021701}{{\em
  Phys. Rev.} {\bf D77} (2008) 021701}
  [\href{http://arxiv.org/abs/0709.2876}{{\tt arXiv:0709.2876}}].

\bibitem{Stahlhofen:2012zx}
M.~Stahlhofen, ``{NLL resummation for the static potential in $\mathcal N = 4$
  SYM theory}'', \href{http://dx.doi.org/10.1007/JHEP11(2012)155}{{\em JHEP}
  {\bf 1211} (2012) 155} [\href{http://arxiv.org/abs/1209.2122}{{\tt
  arXiv:1209.2122}}].

\bibitem{Prausa:2013qva}
M.~Prausa and M.~Steinhauser, ``{Two-loop static potential in $\mathcal N = 4$
  supersymmetric Yang-Mills theory}'',
  \href{http://dx.doi.org/10.1103/PhysRevD.88.025029}{{\em Phys. Rev.} {\bf
  D88} (2013) 025029} [\href{http://arxiv.org/abs/1306.5566}{{\tt
  arXiv:1306.5566}}].

\bibitem{Rey:1998ik}
S.-J. Rey and J.-T. Yee, ``{Macroscopic strings as heavy quarks in large N
  gauge theory and anti-de Sitter supergravity}'',
  \href{http://dx.doi.org/10.1007/s100520100799}{{\em Eur. Phys. J.} {\bf C22}
  (2001) 379--394} [\href{http://arxiv.org/abs/hep-th/9803001}{{\tt
  hep-th/9803001}}].

\bibitem{Maldacena:1998im}
J.~M. Maldacena, ``{Wilson loops in large N field theories}'',
  \href{http://dx.doi.org/10.1103/PhysRevLett.80.4859}{{\em Phys. Rev. Lett.}
  {\bf 80} (1998) 4859--4862} [\href{http://arxiv.org/abs/hep-th/9803002}{{\tt
  hep-th/9803002}}].

\bibitem{Beem:2013hha}
C.~Beem, L.~Rastelli, A.~Sen and B.~C. van Rees, ``{Resummation and S-duality
  in $\mathcal N = 4$ SYM}'',
  \href{http://dx.doi.org/10.1007/JHEP04(2014)122}{{\em JHEP} {\bf 1404} (2014)
  122} [\href{http://arxiv.org/abs/1306.3228}{{\tt arXiv:1306.3228}}].

\bibitem{Beem:2013qxa}
C.~Beem, L.~Rastelli and B.~C. van Rees, ``{$\mathcal N = 4$ Superconformal
  Bootstrap}'', \href{http://dx.doi.org/10.1103/PhysRevLett.111.071601}{{\em
  Phys. Rev. Lett.} {\bf 111} (2013) 071601}
  [\href{http://arxiv.org/abs/1304.1803}{{\tt arXiv:1304.1803}}].

\end{thebibliography}\endgroup
\end{document}